\documentclass[prd,preprintnumbers,longbibliography,
superscriptaddress,tightenlines,twocolumn,nofootinbib
]{revtex4-1}

\usepackage[utf8]{inputenc}
\usepackage{slashed}
\usepackage{amsmath}
\usepackage{amsfonts}
\usepackage{dsfont}
\usepackage{xcolor}
\usepackage{tikz}
\usepackage{hyperref}
\usepackage{color}
\usepackage{orcidlink}
\usepackage{xspace}
\usepackage{placeins}
\usepackage{braket}
\usepackage{cleveref}

\allowdisplaybreaks
\usepackage{ulem}

\usetikzlibrary{decorations.shapes,shapes.geometric}

\newcommand{\dd}{\mathrm{d}}
\newcommand{\psibar}{\bar{\psi}}
\newcommand{\renyi}{R\'enyi}
\newcommand{\GS}{\left |\Omega \right\rangle}
\newcommand{\braGS}{\left \langle\Omega \right|}

\begin{document}

\title{Entanglement asymmetry in gauge theories: chiral anomaly in the finite temperature massless Schwinger model}

\author{Adrien Florio \,\orcidlink{0000-0002-7276-4515}}
\email[]{aflorio@physik.uni-bielefeld.de}
\affiliation{Fakultät für Physik, Universität Bielefeld, D-33615 Bielefeld, Germany}

\author{Sara Murciano \,\orcidlink{0000-0002-1638-5692}}\email{sara.murciano@universite-paris-saclay.fr}
\affiliation{Universit\'e Paris-Saclay, CNRS, LPTMS, 91405, Orsay, France}

\date{\today}

\begin{abstract}
The entanglement asymmetry has emerged in recent years as a practical quantity to study phases of matter.
We present the first study of entanglement asymmetry in gauge theories by considering the chiral anomaly of the analytically solvable massless Schwinger model at both  zero and finite temperatures.  At zero temperature, we find the asymmetry exhibits logarithmic growth with system size. At finite temperature, we show that it is parametrically more sensitive to chiral symmetry-breaking than the corresponding local order parameter: while the chiral condensate decays exponentially, the asymmetry decreases only logarithmically. This establishes the entanglement asymmetry as a promising tool to probe (finite-temperature) phase transitions in gauge theories.
\end{abstract}


\maketitle

\textit{Introduction.} The understanding of phases of matter is constantly evolving. The original Landau paradigm classifying gapped and gapless matter according to their symmetries and their symmetry-breaking pattern is now understood in terms of generalized symmetries. As such, it encompasses, for example, topological phases of matter that lack conventional local order parameters as well as the symmetry origin of the masslessness of photons, see \cite{Cordova:2022ruw,McGreevy:2022oyu,Brennan:2023mmt,Gomes:2023ahz,Shao:2023gho,Schafer-Nameki:2023jdn,Bhardwaj:2023kri,Iqbal:2024pee} for reviews. 

While progress has been made at the conceptual level, the determination of the nature of the phases of given theories remains a challenge. A particularly acute example is Quantum-Chromodynamics (QCD), the non-Abelian gauge theory describing the strong interaction of the Standard Model of particle physics. The deconfinement transition from massless quarks to massive composite particles (hadrons) remains to be understood in terms of symmetries. In the presence of a net density of baryons (for instance, neutrons, such as in a neutron star), a majority of the phase diagram remains elusive, see \cite{Fukushima:2010bq} for a review. 

In parallel, efforts have been made to devise non-local yet practical order parameters to address these difficulties. Building on information-theoretic quantities from resource theory~\cite{gour2009measuring,vaccaro2008tradeoff} and on the intuition that entanglement behaves differently across phases, ref.~\cite{Ares:2022koq} introduced the ``entanglement asymmetry'' as an efficient diagnostic of symmetry-breaking in many-body quantum systems. In the presence of a symmetry, the reduced density matrix can be decomposed into symmetry sectors, allowing us to resolve the contribution of each sector to the entanglement entropy. When the charge operator does not correspond to a symmetry of the state, attempting such a decomposition quantifies the degree of symmetry breaking via the entanglement asymmetry.

Given that the comprehension of entanglement in quantum field theory has rapidly evolved, starting from the pioneering works on the area law of entanglement \cite{Srednicki:1993im,bombelli1986}, the entanglement in conformal field theories  \cite{Calabrese:2004eu,holzhey1994} and in holography, \cite{Ryu:2006bv} to a growing number of more applied topics including but not limited to the role of entanglement generation in high-energy scattering processes and pair production \cite{Cervera-Lierta:2017tdt, Kharzeev:2017qzs,Afik:2020onf,Florio:2021xvj,Kharzeev:2021yyf,Gong:2021bcp,deJong:2021wsd,Dunne:2022zlx,Florio:2023dke,Belyansky:2023rgh,Barata:2023jgd,Grieninger:2023ehb,Grieninger:2023pyb,Lee:2023urk,Kirchner:2023dvg, Robin:2025ymq}, its impact on the construction of effective field theories \cite{Beane:2018oxh,Klco:2020rga,Klco:2021biu,Klco:2021cxq} and how it contributes to thermalization and hydrodynamization \cite{Kharzeev:2005iz,Berges:2017hne,Zhou:2021kdl,Mueller:2021gxd,Ebner:2023ixq,Yao:2023pht, Grieninger:2023ufa, Ciavarella:2025zqf, Turro:2025sec, Janik:2025bbz, Shao:2025ygy,Ebner:2025pdm,Artiaco:2025qqq}, understanding the behavior of the entanglement asymmetry in more complicated theories, such as gauge theories, promises both conceptual and practical progress.
We take the first steps in this direction by studying the breaking of the chiral charge in the massless Schwinger model \cite{Schwinger:1962tp}, arguably the simplest gauge theory and a model that can be solved analytically. We first perform the computation in the zero-temperature ground state, and we find a logarithmic dependence on the system size.
We also generalize the computation to the finite-temperature Gibbs state. At high temperature, we find the asymmetry to be parametrically more sensitive than the chiral condensate, a local order parameter for the breaking of the chiral symmetry.
Consequently, and similarly to quantum phase transitions, it promises to be a generic tool that does not rely on a local order parameter. 
Together, these results pave the way to more extensive analytical and numerical analyses in gauge theories in $1+1$ and higher dimensions.

\textit{The Schwinger model and its chiral charge.}
We study the massless Schwinger model, the theory that describes the physics of a massless Dirac fermion coupled to a $U(1)$ gauge field in $1+1$ dimensions
\begin{align}\label{eq:lagrangian}
     \mathcal{L} = \psibar(i\gamma^\mu\partial_\mu - g\gamma^\mu A_\mu)\psi - \frac14 F_{\mu\nu}F^{\mu\nu} \ .
\end{align}
Here $\psi$ is a two-component Dirac spinor, $\psibar=\psi^{\dagger}\gamma^0$, where $\gamma^{\mu}$ are $2\times 2$ gamma matrices, $A_{\mu}$ is a $U(1)$ gauge field, $F_{\mu\nu}$ is its field strength tensor, $F_{\mu\nu}=\partial_\mu A_\nu-\partial_\nu A_\mu$ and the gauge coupling $g$ is the electric charge.
The classical action is invariant under both a vector $U(1)$ symmetry $\psi\to e^{i\alpha} \psi$ and a chiral $U(1)$ symmetry $\psi\to e^{i\gamma_5\alpha} \psi$. However, at the quantum level, the chiral symmetry is lost. One way to see this is by observing that the vector current $j_\mu=\psibar\gamma^\mu\psi$ is conserved 
\begin{equation}
\partial^\mu j_\mu=0 \ , 
\end{equation}
while the axial current $j_\mu^5=\psibar\gamma^5\gamma^\mu\psi$ is related to the electric field as \cite{Schwinger:1962tp}
\begin{equation}\label{eq:quantum1}
\partial^\mu j_\mu^5=\frac{g}{2\pi}\epsilon_{\mu\nu} F^{\mu\nu} \ . 
\end{equation}
This is an apparent violation of Noether's theorem and is known as the chiral anomaly.
We can write an equivalent bosonic theory to describe the dynamics of these currents (``bosonization'' is an exact duality in $1+1$ dimensions). Taking into account that the gauge field has no propagating degrees of freedom and can be eliminated by using Gauss law $\partial_1 F^{10} = gj^0$, and using the bosonization dictionary $j_\mu \leftrightarrow \epsilon_{\mu\nu}\partial^\nu \phi, \ j_\mu^5 \leftrightarrow \partial_\mu \phi$, the massless Schwinger model is equivalent to a massive free boson $\phi$ of mass $m=\frac{g}{\sqrt{\pi}}$ \cite{Coleman:1976uz}
\begin{align}
     \mathcal{L} = \frac{1}{2}\partial_{\mu}\phi\partial^{\mu}\phi - \frac{1}{2}m^2\phi^2 \ .
\end{align}
When computing the asymmetry, we will use the Hamiltonian lattice discretization of the free boson. In momentum space, it takes the form
\begin{align}
    \mathcal{H}=\sum_{l} \left( \frac{1}{2}\Pi_l \Pi_{-l} +\frac{\omega(l)^2}{2}\Phi_l\Phi_{-l}\right),
\end{align}
with $\omega(l)^2=m^2+ \left(2\pi l / L\right)^2$ the finite volume dispersion relation and $L$ the total system size. In this convention, the fields have units of $[\Phi_l]\sim\sqrt{L}, [\Pi_l]\sim1/\sqrt{L}$ and the operators are canonical conjugates $[\Phi_l,\Pi_m]=i\delta_{l,m}$. In particular, the chiral charge becomes
\begin{align}
Q_5 = \int \mathrm{d}x \pi(x) \to \sqrt{L} \Pi_0 \ . 
\end{align}
The fact that an \textit{a priori} discrete fermionic charge is mapped onto a continuous bosonic one may seem puzzling. The naive fermionic chiral charge needs to be regularized. This mapping corresponds to a specific regularization choice, see \cite{Manton:1985jm} for a detailed explanation.

\textit{Entanglement asymmetry.}
Another central notion to our work is the entanglement asymmetry. It was originally introduced in the context of quantum resource theory: asymmetric states are thought of as a resource to access the effect of operations that are not invariant under the symmetry~\cite{gour2009measuring,vaccaro2008tradeoff,chitambar2019quantum,aditya2025mpemba,summer2025}. In this sense, the concept has its origins in information theory. However, it has also been used to assess symmetry breaking arising from explicit symmetry-breaking terms, for instance, in many-body Hamiltonians or random circuits~\cite{capizzi2023entanglement,fossati2024entanglement,capizzi2024universal,chen2023, ares2024entanglement, russotto2024non, lastres2024entanglement, fossati2024,kusuki2024,chen2024,russotto2025,Mazzoni:2025otu,lamas2025,Benini:2025lav,ahmad2025}. For concreteness, we specialize to a $U(1)$ symmetry, for which the entanglement asymmetry is defined as 
\begin{equation}\label{eq:asydef}
    \Delta S=S(\rho_{\rm sym})-S(\rho),
\end{equation}
where $S(\rho)=-\mathrm{Tr}[\rho\log(\rho)]$ is the von Neumann entropy, $\rho$ is the density matrix of the system and $\rho_{\rm sym}$ is its symmetrized version. Despite originally being introduced to describe the asymmetry of the full system, its definition has also been extended to study the symmetry breaking or the symmetry restoration within a subsystem, and in this case, $\rho$ is replaced by the reduced density matrix restricted to the subsystem~\cite{Ares:2022koq}. For instance, if we consider a bipartition $A\cup B$, the reduced density matrix over $A$ can be computed as $\rho_A=\mathrm{Tr}_B\rho$, where we have performed a partial trace over the degrees of freedom inside $B$. This framework has enabled the investigation of intriguing anomalous relaxation behaviors, such as a quantum analogues of the Mpemba effect. For a comprehensive overview, we refer the reader to refs.~\cite{review,teza2025}.
In this work, we restrict ourselves to a setup in which $A$ describes the full system.

A further extension of \cref{eq:asydef}, which is particularly convenient both for performing analytical computations and for experimental investigations, is the \renyi\ entanglement asymmetry, a replica version of \cref{eq:asydef},
\begin{equation}\label{eq:asydefn}
   \Delta S^{(n)}=S^{(n)}(\rho_{\rm sym})-S^{(n)}(\rho), 
\end{equation}
where $S^{(n)}(\rho)=1/(1-n)\log \mathrm{Tr}\rho^n$ is the \renyi\ entropy. The replica limit $n\to 1$ yields \cref{eq:asydef}.

\Cref{eq:asydef,eq:asydefn} will be our probe of the chiral anomaly in the massless Schwinger model. Unlike prior studies of asymmetry, \cref{eq:lagrangian} contains no explicit chiral symmetry breaking terms; our analysis studies the asymmetry in a setting where symmetry breaking arises intrinsically at a quantum level rather than from external sources.

\textit{Zero-temperature.} We start with the zero-temperature computation, which will serve as a reference point for the finite-temperature case and potentially future studies of the massive Schwinger model. We aim to compute the asymmetry associated with the chiral charge in the ground state $\GS$ of the Schwinger model. To this end, we need to compute the von Neumann entanglement entropy of the symmetrized density matrix
\begin{equation}
    \rho_5 = \int_{-\pi}^\pi \frac{\mathrm{d}\alpha}{2\pi} e^{-i\alpha Q_5} \GS\braGS e^{i\alpha Q_5} \ . \label{eq:rho5_0}
\end{equation} 
Since the total system is in a pure state, $S(\GS\braGS)=0$.
In terms of the lattice operators introduced above, $Q_5 = \sqrt{L} \Pi_0 $. In the absence of the anomaly, i.e. when the charge is conserved, $Q_5$ has a discrete spectrum. This justifies the use of a discrete projector (with integration limits $\alpha \in [-\pi,\pi]$) in \cref{eq:rho5_0}.

Using the replica trick, the starting point is the trace of the $n$-th power of the symmetrized density matrix 
\begin{equation}
    \mathrm{Tr}(\rho_5^n)=\int_{-\pi}^{\pi}\frac{d\alpha_1d\alpha_2\cdots d\alpha_n}{(2\pi)^n}Z_n(\Vec{\alpha}), \label{eq:rho5n_0}
\end{equation}
where $Z_n(\Vec{\alpha})=\prod_{j=1}^n\langle \Omega |e^{i\alpha_{j,j+1}Q_5}|\Omega \rangle$, $\alpha_{j,j+1}=\alpha_j-\alpha_{j+1}$ and $\alpha_{j+n}\equiv \alpha_j$. The quantity $Z_n(\Vec{\alpha})$ is called the charged moment and, in this case, its computation reduces to the knowledge of the two-point correlator
\begin{equation}
    \langle \Omega |e^{i\alpha \sqrt{L} \Pi_0}|\Omega \rangle.
\end{equation}
Inserting a resolution of the identity in the momentum basis, we can rewrite the correlator above as\footnote{To be fully consistent with the discrete projection, we should in principle restrict ourselves to summing over a discrete subset of momenta. As this does not matter in the thermodynamics limit, we ignore here this subtelty, see also appendix~\ref{app:finitesize} for more details.}
\begin{equation}\label{eq:Gaussintegral}
 \int_{-\infty}^{\infty} dp \langle \Omega |e^{i\alpha \sqrt{L} \Pi_0}|p\rangle\langle p|\Omega \rangle.   
\end{equation}
At this point, the mapping of the massless Schwinger model to a massive bosonic field gives 
\begin{align}
   e^{i\alpha \sqrt{L} \Pi_0}|p\rangle=& \,e^{i\alpha \sqrt{L} p}|p\rangle, \nonumber\\  \langle p|\Omega \rangle=& \,\frac{e^{-p^2/(2m)}}{(m\pi)^{1/4}},
\end{align}
where we have simply used that $\Pi_0|p\rangle=p|p\rangle$.
Plugging these relations into the integral \eqref{eq:Gaussintegral}  yields 
\begin{equation}
   \langle \Omega |e^{i\alpha \sqrt{L} \Pi_0}|\Omega \rangle=e^{-\alpha^2 mL/4}. 
\end{equation}
Therefore, we have to solve the $n$-fold integral 
\begin{equation}\label{eq:charged1}
    \int_{-\pi}^{\pi}\frac{d\alpha_1d\alpha_2\cdots d\alpha_n}{(2\pi)^n}\prod_{j=1}^n e^{-mL/4 \alpha^2_{j,j+1}}.
\end{equation}
By doing a change of variables $\alpha_{j,j+1}\to \gamma_j/\sqrt{L}$ and using $\sum_j \alpha_{j,j+1}=0$, we obtain in the large $L$-limit 
\begin{equation}
    \int_{-\infty}^{\infty}\frac{d\gamma_1d\gamma_2\cdots d\gamma_n}{(2\pi)^{n-1} L^{n/2}}\prod_{j=1}^n e^{-m/4 \gamma^2_{j,j+1}}\delta\left(\sum_i\gamma_i/\sqrt{L}\right). \label{eq:0Tintermediatesteps}
\end{equation}
Using the integral representation of the $\delta$-function leads to
\begin{equation}
 \mathrm{Tr}(\rho_5^n) =\sum_{k=-\infty}^{\infty}J_k^n,\qquad J_k=\int_{-\infty}^{\infty} d\gamma \frac{e^{i\frac{k}{\sqrt{L}}\gamma-\gamma^2 m/4}}{2\pi \sqrt{L}}.
\end{equation}
Evaluating the Gaussian integral and taking the replica limit $n\to 1$, the entanglement asymmetry behaves as 
\begin{equation}\label{eq:asy0T}
    \Delta S=-\sum_k\mathrm{Re}[J_k \log J_k]\simeq \frac{1}{2}\log(mL\pi)+\frac{1}{2},
\end{equation}
where, in the last step, we have approximated the discrete sum over
$k$ by a continuum integral. 

We see that the entanglement asymmetry shows the typical logarithmic scaling $1/2\log L$, which has also been found in different contexts like Gaussian~\cite{murciano2024}, Haar random~\cite{ares2024entanglement}, and generic translation-invariant matrix product states where the symmetry is explictly broken~\cite{capizzi2024universal}. Another interesting feature is that the ``length-scale'' of the problem is $m L$, and strictly speaking, if we take the limit $m\to 0$, the asymmetry diverges. However, if $m\to0$, \cref{eq:quantum1} implies that the chiral symmetry is restored, so we expect the asymmetry to vanish. The reason for this apparent discrepancy is the non-commutation of the limits $m\to 0$ and $L\to\infty$, already observed in \cite{Ares:2022koq}.

The fact that the scaling of the asymmetry for the chiral anomaly matches other very different cases where the symmetry is explicitly broken~\cite{Mazzoni:2025otu} can be traced back to the fact that the ground state of the Schwinger model satisfies the cluster property, i.e., the variance of the operator $Q_5$ grows extensively in the system size, since $\langle Q_5^2\rangle=m L/2$. Therefore, the chiral symmetry-breaking is bounded by \cite{Mazzoni:2025otu} 
\begin{equation}
    \Delta S\leq  \frac{1}{2}\log\left[2\pi \left(\frac{mL}{2}+\frac{1}{12}\right)\right]+\frac{1}{2},
\end{equation}
which is consistent with our result.

\textit{Finite temperature.} Having the zero-temperature case as a reference, we can now study what happens at finite temperature. While the massless Schwinger model does not exhibit a phase transition at any finite temperature, we expect on general grounds the symmetry to be progressively restored as the temperature increases. Indeed, in the limit where the inverse temperature $\beta$ goes to zero, the Gibbs ensemble tends to the (normalized) Identity, which is symmetric with respect to any arbitrary charge. This is an ideal setting to probe the behavior of the entanglement asymmetry: we expect it to get smaller as the temperature goes up. In this case, we also explicitly know the relevant order parameter that detects chiral symmetry-breaking. This will allow us to compare their temperature dependence. 

To do so, we replace in \cref{eq:rho5n_0} the ground state $|\Omega\rangle\langle\Omega|$ density matrix by the Gibbs state $e^{-\beta H}/Z$ 
\begin{align}
\mathrm{Tr}\left(\rho_5^n(\beta)\right)=\int_{-\pi}^{\pi}&\frac{d\alpha_1d\alpha_2\cdots d\alpha_n}{(2\pi)^n}\label{eq:SnT}
\\
&\cdot\mathrm{Tr}\left(\prod_{j=1}^n e^{-\alpha_{j,j+1}Q_5}e^{-\beta H}\right) \ \notag \ .
\end{align}
Once again, the expression above can be rewritten in a more convenient form working in the momentum basis. Since the steps are rather technical, we report them in appendix \ref{app:computation} and move directly to the main results.

The expression of the charged moments to be plugged in \cref{eq:SnT} is
\begin{align}\label{eq:ratio}
    \frac{\mathrm{Tr}(\rho_5^n(\beta))}{\mathrm{Tr}(\rho^n(\beta))} \notag
&=\int_{-\pi}^{\pi}\frac{d\alpha_1d\alpha_2\cdots d\alpha_n}{(2\pi)^n}\exp\left( -\frac12 \vec A^T \mathbf{M}^{-1} \vec A L \right),
\end{align}
where
\begin{align}
    A_i &= \alpha_{i,i+1} \ , \
    M_{ii} = \frac{2\coth(m\beta)}{m}\\
    M_{i,i+1} &= M_{i-1,i} = -\frac{1}{m\sinh(m\beta)} (1+\delta_{n2})\ , \label{eq:matrixel}
\end{align}
with the identification $0\equiv n$ and $n+1\equiv 1$.
After computing explicitly the integral above, we get
\begin{align}
    \frac{\mathrm{Tr}(\rho_5^n(\beta))}{\mathrm{Tr}(\rho^n(\beta))} &=\frac{2 \sinh\left(\frac{\beta m n}{2}\right)}{(2\pi L m\sinh(m\beta))^{n/2}}\nonumber\\
    &\sum_{l=-\infty}^{\infty}\exp\left(-\frac{nl^2}{mL} \tanh\left(\frac{\beta m}{2}\right)\right),
    \label{eq:rho5ovtr}
\end{align} 
which leads to the following expression for the entanglement asymmetry
\allowdisplaybreaks[0]
\begin{align}\label{eq:asymm_finiteT}
    \Delta S=&\frac{\sinh(m\beta/2)}{\sqrt{2\pi m L \sinh(m\beta)}} \\
    &\cdot \Big( - m\beta\coth(m\beta/2)\theta_3\left(e^{-\tanh(m\beta/2)/(mL)}\right)\notag\\
    &+\theta_3\left(e^{-\tanh(m\beta/2)/(mL)}\right)\log\left(2\pi m L \sinh(m\beta)\right)\notag\\
    &+\frac{2\tanh(m\beta/2)}{mL}\sum_l l^2e^{-l^2\tanh(m\beta/2)/(mL)} \Big )\ \ .\notag
\end{align}
\allowdisplaybreaks
Here $\theta_3$ denotes the Jacobi theta function. 
Even though this expression is rather involved, it admits simple asymptotes in two regimes. The first one is the zero temperature limit, $m\beta\gg 1$, where \cref{eq:asymm_finiteT} reduces to 
\begin{equation}
    \Delta S\approx \frac{1}{2}(1+\log(mL\pi)). 
\end{equation}
This is a good cross-check since we recover the result \eqref{eq:asy0T}. The other interesting limit is the high-temperature regime, which is valid as far as  $1/(mL)<m\beta\ll mL$, and reads
\begin{equation}
   \Delta S\approx \frac{1}{2}(-1+\log(2\pi m^2 \beta L)). 
   \label{eq:smallBeta}
\end{equation}
As in the zero-temperature case, we can also verify the upper bounds on the asymmetry coming from clustering. We compute explicitly the variance of $Q_5$ at finite temperature $\langle Q_5^2\rangle_{\beta}=\mathrm{Tr}(e^{-\beta H} Q_5^2)/Z$, and  find that 
\begin{equation}
  \langle Q_5^2\rangle_{\beta}=  \frac{mL }{2}\coth\left(\frac{\beta m}{2}\right).
\end{equation}
If we use the expression for the entanglement asymmetry in \cref{eq:asymm_finiteT}, we can check that the bound 
\begin{equation}
    \Delta S\leq \frac{1}{2}\log\left[2\pi  \left(\frac{mL }{2}\coth\left(\frac{\beta m}{2}\right)+\frac{1}{12}\right)\right]+\frac{1}{2},
\end{equation}
is satisfied for any value of $\beta$. 

\begin{figure}
\includegraphics[width=0.48\textwidth]{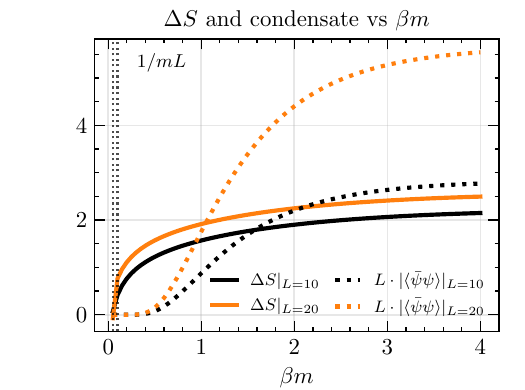}
	\caption{Entanglement asymmetry (solid lines) and rescaled chiral condensate density $\langle\psibar\psi\rangle \cdot L$ (dashed lines) as functions of inverse temperature $\beta$ for different system sizes $L$.  The logarithmic behavior of the asymmetry contrasts with the exponential suppression of the order parameter at high temperature. It also constrasts with the extensive behavior of the order parameter as a function of $mL$. The vertical dotted lines show the values of $1/mL$; we expect the thermodynamics limit to be reliable for $\beta m > 1/mL$.}
	\label{fig:ea}
\end{figure}

\begin{figure}
\includegraphics[width=0.48\textwidth]{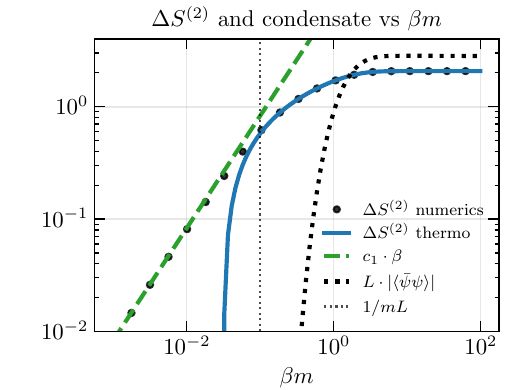}
  \caption{Second \renyi\ asymmetry $\Delta S^{(2)}$ and rescaled chiral condensate density $\langle\psibar\psi\rangle \cdot L$ as functions of $\beta$ for $mL=10$. Black dots show numerical results at finite $L$, while the blue curve shows the thermodynamic limit expression from \cref{eq:rho5ovtr}. The green dashed line indicates the small-$\beta$ asymptotic behavior $\Delta S^{(2)} \sim (\pi^2/12) L m^2 \beta$. The agreement between finite-$L$ and thermodynamic results is excellent for $\beta m > 1/(mL)$.}
	\label{fig:renyi2}
\end{figure}

A natural question to ask now is what we learn about the chiral anomaly from the entanglement asymmetry that we cannot learn from the study of the order parameter. We report here the finite-temperature result of the order parameter found in \cite{Sachs:1991en}
\begin{equation}\label{eq:finiteTLcond}
 \langle\psibar\psi\rangle=-\frac{2}{\beta}e^{-\frac{\pi}{\beta m}\coth\left(\frac{L m}{2}\right)}e^{F(\beta m)-2H(\beta m,L/\beta )},
\end{equation}
where 
\begin{equation}
    \begin{split}
    F(x)&=\sum_{n>0}\left(\frac{1}{n}-\frac{1}{\sqrt{n^2+(x/2\pi)^2}}\right),\\
H(x,\tau)&=\sum_{n>0}\frac{1}{\sqrt{n^2+(x/2\pi)^2}}\frac{1}{e^{\tau\sqrt{(2\pi n)^2+x^2}}-1}.
    \end{split}
\end{equation}
We compare in \cref{fig:ea} the behavior of the order parameter and of the entanglement asymmetry as a function of $\beta$ for sufficiently large $L$. To have dimensionless
quantities, we rescale the chiral condensate density $\psibar\psi$ by the lattice volume $L$. Both quantities detect that the Gibbs state is not symmetric under chiral rotation. The main difference is their scaling with volume and how this interplays with temperature.
As expected, both go to zero at high temperature. However, in the regime $1/(mL)< m\beta <1$, \cref{fig:ea} shows a striking difference: the asymmetry exhibits logarithmic behavior \eqref{eq:smallBeta} while the condensate decays exponentially as $\langle\psibar\psi\rangle\sim e^{-\pi/m\beta}$.
The presence of quantum superpositions between different symmetry sectors captures the symmetry-breaking. This global property of the state is measured only by the asymmetry and not by the local order parameter. It leads to the parametric enhancement of the former over the latter. 
Even though, strictly speaking, we do not know how to compute analytically the asymmetry for $\beta\ll 1/(mL)$, we can, however, extend our analysis to small $m\beta$ by directly computing the \renyi\ asymmetries numerically using \cref{eq:SnT,eq:ratio}. We show the result for the second \renyi\ $\Delta S^{(2)}$ in \cref{fig:renyi2}. Black dots correspond to the numerical computation. The plain blue comes from \cref{eq:rho5ovtr} and is the thermodynamic limit. We first see that both indeed agree very well for $\beta m>1/mL$. As a result, the parametric enhancement is also visible in the \renyi\ asymmetries. Furthermore, it persists in the finite size regime $\beta m<1/mL$, as we can show that there $\Delta S^{(2)}\sim (\pi^2/12) L m^2 \beta$ (green dashed line) is linear in $\beta$ while the condensate \cref{eq:finiteTLcond} still exponentially decays. We further support this behavior for larger \renyi\ indices in appendix \ref{app:finitesize}. This confirms that the striking parametric enhancement of the asymmetry is not an artifact of how we computed the thermodynamic limit.

\textit{Discussion.}  The massless Schwinger model is analytically solvable and, through its chiral anomaly, exhibits a non-trivial symmetry breaking even in $1+1$ dimensions. We showed that this is detected by the entanglement asymmetry. Its dependence is logarithmic in the system size. We were also able to compute its temperature dependence and compare it to that of the chiral condensate, the relevant local order parameter. While both quantities go to zero at infinite temperature, we showed that the asymmetry is parametrically more sensitive to symmetry-breaking at large temperatures $m\beta< 1$. In particular, while the order parameter is exponentially suppressed at high temperature, the \renyi\ entanglement asymmetry only decreases polynomially in $m\beta$. It is reasonable to expect this property to be rather ubiquitous. It is inherited from the logarithmic dependence of the entropy on system size, which is generic for systems obeying the cluster decomposition principle.
This provides strong incentives to develop the asymmetry as a more sensitive tool for studying (finite-temperature) phase transitions. This is particularly true as recent advances have shown that \renyi\ entropies can be extracted from Euclidean Monte-Carlo simulations of realistic theories in higher dimensions \cite{Polikarpov:2008tg,Bulgarelli:2023ofi,Bulgarelli:2024onj, Amorosso:2024leg, Amorosso:2024glf}. Generalizing these methods to compute \renyi\ asymmetries is a natural outlook. 

Beyond static properties, we have already mentioned that the entanglement asymmetry is also useful to
quantify how dynamics can restore a broken symmetry locally and show anomalous symmetry restoration phenomena, like a quantum version of the Mpemba effect in closed systems.  It would be interesting to consider whether this also happens in gauge theories.  A simple thought experiment is to start from the ground state of the massless Schwinger model ($m\neq 0$) and perform a quench with $m=0$. In this case, as the chiral anomaly cannot be distinguished from an explicit symmetry-breaking, we would expect a quantum version of the Mpemba effect to persist. Confirming this and investigating it in realistic finite-temperature phase transitions is an exciting outlook.
More generally, this work opens up new possibilities to study symmetry-breaking or symmetry restoration in Hamiltonian lattice gauge theories. Given the success that Gaussian techniques have had in the study of the entanglement asymmetry, exploiting the Gaussian variational ansatz~\cite{sala2018} could be a first step in this direction.

Another direction more specific to gauge theories is understanding better the exact relation between such entanglement measures and anomalous symmetries in higher dimensions. In $3+1$ dimensions, the breaking of the chiral charge can be understood as one aspect of a conserved non-invertible symmetry \cite{Choi:2022jqy, Cordova:2022ieu}. The presence of a chiral condensate is not automatic, and when present it does not simply amount to an explicit symmetry-breaking as in the Schwinger model, but leads to the existence of a massless Goldstone mode \cite{GarciaEtxebarria:2022jky}.  Recently, refs.~\cite{Benini:2025lav,ahmad2025} initiated the study of the entanglement asymmetry for non-invertible symmetries. Extending this and our work in the context of gauge theories and anomalies, and understanding the entanglement structure of such theories as well as how it rearranges across phases is of great interest.

\textit{Acknowledgment.} The authors thank SwissMAP for funding the ALPS Summer School series, without which this work would not have existed. A.F. acknowledges interesting discussions with  G.~Cuomo and D.~Kharzeev. S.M. acknowledges F.~Ares, P.~Calabrese, G. Di Giulio, and P.~Sala. A.F. is supported by the Deutsche Forschungsgemeinschaft (DFG, German Research Foundation) through the Emmy Noether Programme Project No. 545261797 and the  CRC-TR 211 "Strong-interaction matter under extreme conditions," Project No. 315477589-TRR 211.

\FloatBarrier

\appendix

\section{Analytical derivation of the finite-temperature case}
\label{app:computation}
We present here the computation of the charged moments \eqref{eq:rho5ovtr} used to compute the \renyi\ asymmetries.
Working again in the momentum basis allows us to rewrite this as 
\begin{align}
\mathrm{Tr}\left(\rho_5^n(\beta)\right)=\int_{-\pi}^{\pi}&\frac{d\alpha_1d\alpha_2\cdots d\alpha_n}{(2\pi)^n} Z_n(\vec\alpha,\beta)
\end{align}
with \begin{align}
Z_n(\vec\alpha,\beta)&=\int\dots\int \dd\vec{p}_1\dots\dd \vec{p}_n\notag\\
& \ \ \ \ \ \  \ \ \ \ \ \ \ \ \ \prod_{j=1}^n\braket{\vec{p}_j|e^{i\alpha_{j,j+1}Q_5}\frac{e^{-\beta H}}{Z_1}|\vec{p}_{j+1}}\label{eq:Zbeta1} \\
&\hspace{-1.4cm}= \int\dots\int \dd{p^0}_1\dots\dd {p^0}_n  \prod_{j=1}^n\braket{{p^0}_j|e^{i\alpha_{j,j+1}Q_5}\frac{e^{-\beta H}}{Z_1}|{p^0}_{j+1}} \notag\\
&\hspace{-1.4cm}\cdot \int\dots\int \dd\vec{p'}_1\dots\dd \vec{p'}_n\prod_{j=1}^n\braket{\vec{p'}_j|\frac{e^{-\beta H}}{Z_1}|\vec{p'}_{j+1}}\label{eq:Zbeta2} \\
&\equiv Z^0_n(\vec{\alpha},\beta) \cdot Z'_n(\beta) \ .
\end{align}
As apparent in \cref{eq:Zbeta1}, we insert $n$ times the identity, resolved in the momentum basis $\mathbf{1}=\prod_l\int_{-\infty}^\infty \ket{p^l}\bra{p^l}$. The product over $l$ is a product over lattice sites; we use a vectorial notation $\ket{\vec{p}}=\ket{\dots,p^0,p^1,\dots}$ to make it implicit. In \cref{eq:Zbeta2}, we recognize that the chiral charge acts only on site $0$, and split out this contribution into $Z_n^0(\vec\alpha, \beta)$. We use primes to denote the exclusion of the site $0$. Note also that $Z'(\beta)$ is simply a product of thermal partition functions.  

Using the explicit representation of the propagator of the simple harmonic oscillator in momentum space \cite{Feynman2010-gl}
\begin{widetext}
\begin{equation}
    \braket{q^0|e^{-\beta H}|p^0} = \frac{1}{\sqrt{2\pi m \sinh(\beta m)}} \exp\left(-\frac{\coth(\beta m)}{2 m}\left((p^0)^2+(q^0)^2\right)+\frac{p^0q^0}{m\sinh(\beta m)}\right) \ ,
\end{equation}
\end{widetext}
we can rewrite 
\begin{align}\label{eq:momentum}
    Z^0_n(\vec{\alpha}) &=   \frac{1}{Z(\beta)^n}  \frac{1}{\left( 2\pi m \sinh(\beta m)\right )^{n/2}}\notag\\
    &\int\dots\int \dd{p^0}_1\dots\dd {p^0}_n e^{i\sqrt{L} \vec A \cdot \mathbf{p}^0} e^{-\frac12(\mathbf{p}^0)^T \cdot \mathbf{M} \cdot \mathbf{p}^0},
\end{align}
where $Z(\beta)=2\sinh(m\beta/2)/(m\sinh(m\beta))$ the finite temperature partition function of a  simple harmonic oscillator. As we can see in \cref{eq:matrixel}, in this notation special care needs to be given to the case $n=2$. This however does not affect the final result, which is valid for all $n>1$. The Gaussian integrals can be readily performed to get
\begin{align}
    Z^0_n(\vec{\alpha}) &=  \frac{1}{Z(\beta)^n}  \frac{1}{\left( m \sinh(\beta m)\right )^{n/2}} \frac{1}{\sqrt{\mathrm{det}(M)}} \notag\\
    &\cdot \exp\left( -\frac12 \vec A^T \mathbf{M}^{-1} \vec A L \right) \label{eq:z0nalphav1}\\
    &=  \frac{Z(n\beta)}{Z(\beta)^n} \exp\left( -\frac12 \vec A^T \mathbf{M}^{-1} \vec A L \right) \label{eq:z0nalphav2}
\end{align}
where we compute explicitly $\mathrm{det}(M)$ to go from \cref{eq:z0nalphav1} to \cref{eq:z0nalphav2}. And indeed, \cref{eq:z0nalphav2} reproduces the correct result in the absence of asymmetry $\alpha_i=0$. 

Performing the same manipulations as in \cref{eq:0Tintermediatesteps}, we can proceed with the computation of the moments as 
\begin{align}
    \mathrm{Tr}\left(\rho_5^n(\beta)\right) &=  \frac{Z'_n(\beta)}{L^{n/2}{(2\pi)^n}}  \cdot \label{eq:z0nalphav3} \\
    &\int_{-\infty}^{\infty} \dd\gamma_1 \dots \dd \gamma_n \sum_{l=-\infty}^{\infty}Z^0_n(\vec{\gamma}) \exp\left(\frac{il}{\sqrt{L}}\sum_i \gamma_i\right)\notag \ . 
\end{align}
Swapping the sum and the integral allows us to reduce this again to a multivariate Gaussian integral.  Performing these integrals explicitly leads to \cref{eq:rho5ovtr}. 

\subsection{Alternative derivation of the charged moments}
In this section, we provide an alternative derivation of the charged moments, and we show that we get the same result as in eq. \eqref{eq:charged1}. Rather than working in the momentum basis, we can compute the charged moments in the occupation number basis: each state is labeled by $|\Vec{m}\rangle=|t^0,t^1,\dots,t^{L-1}\rangle$ and
\begin{equation}
Z_n(\vec{\alpha})=\sum_{\vec{t}_1,\dots, \vec{t}_n}\prod_{j=1}^n\braket{\vec{t}_j|e^{i\alpha_{j,j+1}Q_5}\frac{e^{-\beta H}}{Z_1}|\vec{t}_{j+1}}
\label{eq:znalpha}
\end{equation}
The charge $Q_5$ acts non trivially only on the zero mode, so we can isolate the term 
\begin{widetext}
\begin{equation}
\begin{split}
&\braket{t^0_j|e^{i\alpha Q_5}\frac{e^{-\beta H}}{Z_1}|t^0_{j+1}}=\frac{e^{-\beta m t_{j+1}^0}}{Z_1}\braket{t^0_j|e^{i\alpha Q_5}|t^0_{j+1}}\\
&
=\frac{e^{-\beta m t_{j+1}^0}}{Z_1}e^{\alpha^2 mL/4}\sum_{kk'}\frac{(-1)^{k}}{k!k'!}\left(\alpha\sqrt{\frac{m L}{2}}\right)^{k+k'}\braket{t^0_j|(a_0^{\dagger})^k(a_0)^{k'}|t^0_{j+1}}\\&=\frac{e^{-\beta M t_{j+1}^0}}{Z_1}e^{\alpha^2 mL/4}\sum_{kk'}\frac{(-1)^{k}}{k!k'!\sqrt{t_j^0!t^0_{j+1}!}}\left(\alpha\sqrt{\frac{m L}{2}}\right)^{k+k'}\braket{0|(a_0^{\dagger})^{k+t^0_j}(a_0)^{k'+t^0_{j+1}}|0}\\&=\frac{e^{-\beta m t_{j+1}^0}}{Z_1}e^{\alpha^2 mL/4}\sum_{k}\frac{(-1)^{k}(t_j^0+k)!}{k!(k+t^0_j-t^0_{j+1})!\sqrt{t_j^0!t^0_{j+1}!}}\left(\alpha\sqrt{\frac{m L}{2}}\right)^{2k+t_j^0-t^0_{j+1}}\\&=\frac{e^{-\beta m t_{j+1}^0}}{Z_1}e^{\alpha^2 mL/4}\frac{1}{(t^0_j-t^0_{j+1})!}\sqrt{\frac{t_j^0!}{t^0_{j+1}!}}\left(\alpha\sqrt{\frac{m L}{2}}\right)^{t_j^0-t^0_{j+1}} \,_1 F_1(1+t^0_j,1+t_j^0-t^0_{j+1},-mL/2\alpha^2),
\end{split}
\end{equation}
\end{widetext}
where $_1 F_1$ denotes the hypergeometric function.
The action of the non-zero modes gives ($H$ here does not include the zero mode)
\begin{equation}
\begin{split}
    &\sum_{\vec{t}=(t^1,\dots,t^{L-1})}\frac{\braket{\vec{t}|e^{-\beta n H}|\vec{t}}}{Z_1^n}=\\&\frac{1}{Z_1^n}\prod_{l\neq 0}\sum_{t^l=0}^{\infty}e^{-\beta \omega_l n t^l}=\prod_{l\neq 0}\frac{(1-e^{-\beta \omega_l})^n}{1-e^{-\beta \omega_l n}}.
\end{split}
\end{equation}
Putting everything together, we get 
\begin{equation}\label{eq:Znalpha}
\begin{split}
   &Z_n(\vec{\alpha})=(1-e^{-\beta m})^n\prod_{l\neq 0}\frac{(1-e^{-\beta \omega_l})^n}{1-e^{-\beta \omega_l n}}\\&\prod_{j=1}^n\sum_{t_j} e^{-\beta m t^0_j}e^{\alpha^2_{j,j+1}mL/4}\left(\alpha_{j,j+1}\sqrt{\frac{mL}{2}}\right)^{t_j^0-t_{j+1}^0}\\&\times\sqrt{\frac{t_j^0!}{t^0_{j+1}!}} \frac{1}{(t_j^0-t_{j+1}^0)!}\,\\&_1 F_1(1+t^0_j,1+t_j^0-t^0_{j+1},-mL/2\alpha_{j,j+1}^2),
\end{split}
\end{equation}
with the identification $t^0_{n+1}=t^0_1$. 

We can use the property of the hypergeometric functions 
\begin{equation}
    \,_1F_1(1+a,1+a-b,-x)=e^{-x}\,_1F_1(-b,1+a-b,x),
\end{equation}
and relate them to the Laguerre polynomial as
\begin{equation}
    \,_1F_1(-\ell,1+k,x)=\frac{k! \ell!}{(k+\ell)!}L^{(k)}_{\ell}(x).
\end{equation}
We keep in mind that, if $k<0$, we perform a shift $L_n^{(-m)}(x)=(-x)^m\frac{(n-m)!}{n!}L_{n-m}^{(m)}(x)$. 
At the end, we get
\begin{equation}
\begin{split}
   \frac{ Z_n(\vec{\alpha})}{Z_n(\vec{0})}=&\frac{1}{\mathcal{N}_n}\prod_{j=1}^n\sum_{t_j} e^{-\beta m t_j}e^{-\alpha^2_{j,j+1}mL/4}\\&\left(\alpha_{j,j+1}\right)^{t_j-t_{j+1}} L^{(t_j-t_{j+1})}_{t_{j+1}}(\alpha^2_{j,j+1}mL/2),
\end{split}
\end{equation}
where 
\begin{equation}
    \mathcal{N}_n=\prod_{j=1}^n\sum_{t_j} e^{-\beta m t_j}=\frac{1}{(1-e^{-\beta m})^n}.
\end{equation}
Using the integral identities between the generalized Lagurre and Hermite polynomials, we can rewrite the quantity above as 
\begin{equation}\label{eq:ratio1}
\begin{split}
&\frac{Z_n(\vec{\alpha})}{Z_n(\vec{0})}=\frac{1}{\mathcal{N}_n}\prod_j \sum_{t_j}\displaystyle\int_{-\infty}^{\infty}dp_je^{-\beta m t_j}e^{i\alpha_{j,j+1}p_j\sqrt{L}}e^{-p_j^2/m}\\&\frac{1}{\sqrt{2^{t_j+t_{j+1}}t_j!t_{j+1}!}}H_{t_j}(p_j/\sqrt{m})H_{t_{j+1}}(p_j/\sqrt{m}).
\end{split}
\end{equation}
We can reshuffle all the terms to perform first a sum over $t_j$, which reads
\begin{equation}
    \sum_{t_j}\frac{e^{-\beta m t_j}}{2^{t_j}t_j!}H_{t_j}(t_j/\sqrt{m})H_{t_j}(p_{j+1}/\sqrt{m})e^{-p_j^2/(2m)}e^{-p_{j+1}^2/(2m)}.
\end{equation}
By applying the Mehler‐kernel identity, the sum above can be rewritten as (neglecting $n,\alpha$-independent factors, as they will be cancelled by the denominator $Z_n(\vec{0})$)
\begin{equation}
   \sqrt{\frac{e^{\beta m}}{2\pi\sinh(\beta m)}} e^{-\coth(\beta m)/(2m)(p_j^2+p_{j+1}^2)+p_j p_{j+1}/(m \sinh(\beta m))}.
\end{equation}
If we plug this into eq. \eqref{eq:ratio1}, we get 
\begin{equation}
\begin{split}
 &\frac{Z_n(\vec{\alpha})}{Z_n(\vec{0})}=
\frac{1}{\mathcal{N'}_n}  \displaystyle\int_{-\infty}^{\infty}\prod_jdp_j \\&e^{\sum_j [i\alpha_{j,j+1}p_j\sqrt{L}-\coth(\beta m)/mp_j^2 +p_j p_{j+1}/(m \sinh(\beta m))]},
 \end{split} 
\end{equation}
where 
\begin{equation}
 \mathcal{N'}_n=    \displaystyle\int_{-\infty}^{\infty}\prod_jdp_j e^{\sum_j[ -\coth(\beta m)/mp_j^2 +p_j p_{j+1}/(m \sinh(\beta m))]}.
\end{equation}

This result matches eq. \eqref{eq:momentum} for the ratio between the charged and the uncharged moments.

\section{Finite-size \renyi\ entropies}\label{app:finitesize}

We present in this appendix some results about the behavior of the \renyi\ entropies over the whole $\beta m$ range.  As mentioned in the main text, \cref{eq:SnT} allows for a direct numerical computation of the \renyi\ entropies for any value of $m\beta$ at finite $mL$. 

We need to face here an ambiguity that was already mentioned in the main text. While the conserved charged is discrete, leading to the use of discrete projectors in \cref{eq:SnT} (i.e. a compact integration between $-\pi$ and $\pi$), the regularized anomalously-non-conserved charge is continuous, reflected in the continuous spectrum of the momentum operator. This introduces an ambiguity that is irrelevant as $mL\to\infty$ but that matters at  small $mL$: the charged moment $Z_n(\vec \alpha,\beta)$ is not strictly periodic anymore  in $\alpha_i-\alpha_{i+1}$. This is a consequence of the continuous integration over momenta mentioned in the main text. One solution is to restrict the continuous spectrum to a discrete subset $p=n/\sqrt{L}$. Another is to replace \cref{eq:SnT} by the discrete equivalent of \cref{eq:z0nalphav3}
\begin{align}\label{eq:z0nalphav4}
\mathrm{Tr}\left(\rho_5^n(\beta)\right) &=  \frac{Z'_n(\beta)}{L^{n/2}{(2\pi)^n}}  \cdot  \\
&\int_{-\pi}^{\pi} \dd\gamma_1 \dots \dd \gamma_n \sum_{l=-\infty}^{\infty}Z^0_n(\vec{\gamma}) \exp\left(\frac{il}{\sqrt{L}}\sum_i \gamma_i\right)\notag \ . 
\end{align}
\Cref{eq:SnT}  and \cref{eq:z0nalphav4} are equivalent  when $Z_n(\vec \alpha,\beta)$ is periodic but  \cref{eq:z0nalphav4} enforces it when it's not.  We checked that both solution leads to virtually identical results for $mL>1$; the finite $m L$ numerical results presented in this work were obtained with the second method. 

\begin{figure}
\includegraphics[width=0.48\textwidth]{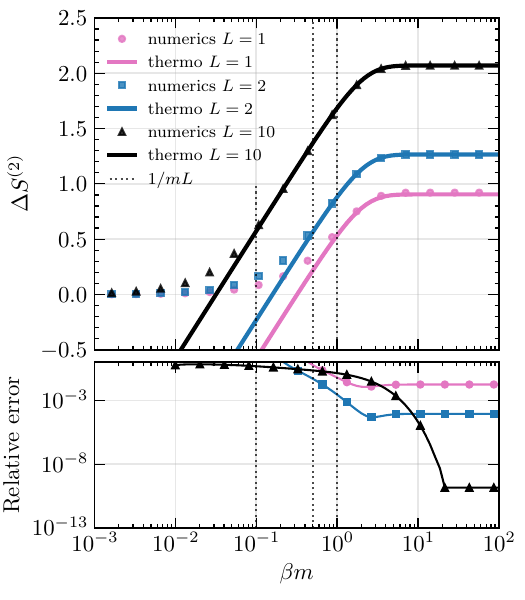}
	\caption{Second \renyi\ asymmetry $\Delta S^{(2)}$ for different system sizes: numerical results at finite $mL$ (markers) versus thermodynamic limit analytics (solid curves) in the upper panel, and relative error in the lower panel. The results show rapid convergence as $mL$ increases. Colors indicate $mL=1$ (pink), $mL=2$ (blue), and $mL=10$ (black).}
	\label{fig:finiteLComp}
\end{figure}

Computing numerically \eqref{eq:z0nalphav4}  allows us to first check in more detail the convergence to the thermodynamic limit in \cref{fig:finiteLComp}. The upper plot shows the behavior of the second \renyi\ $\Delta S^{(2)}$ for $mL=1,2,10$ (pink, blue, black markers) and their thermodynamic limit (pink, blue, black  plain curves). The lower plot displays the relative errors. They reduce extremely fast as $L$ grows as long as $\beta m>1/mL$. The case $mL=10$ is even dominated by the errors coming from the numerical integration.

\begin{figure}
\includegraphics[width=0.48\textwidth]{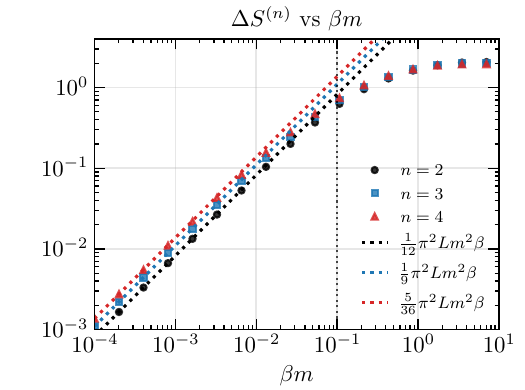}
	\caption{\renyi\ entanglement asymmetries $\Delta S^{(n)}$ for $n=2,3,4$ as functions of $\beta$ for $mL=10$. Solid curves show numerical results, while dashed lines indicate the small-$\beta$ asymptotic behaviors, all of which are linear in $\beta$.}
	\label{fig:nthrenyi}
\end{figure}

We study the dependence of the $n$-th \renyi\ entanglement asymmetry on $n$ using the same strategy. Explicitly computing the matrix $\mathbf{M}$ (\cref{eq:matrixel}) at finite $n$ allows us to compute \cref{eq:rho5n_0} explicitly and extract the following $\beta m \to 0$ asymptotes
\begin{align}
\Delta S^{(2)} &\sim \frac{\pi^2}{12} L m^2 \cdot \beta \\  
\Delta S^{(3)} &\sim \frac{\pi^2}{9} L m^2 \cdot \beta\\  
\Delta S^{(4)} &\sim \frac{5 \pi^2}{36} L m^2 \cdot \beta \ .
\end{align}
Their full dependence on $m\beta$ is displayed in \cref{fig:nthrenyi}.
The \renyi\ entropies are linear in $\beta$ at small $\beta$, in contrast to the quadratic $\beta$-dependence found for free fermions at finite-temperature~\cite{ares2024}.

\FloatBarrier

\clearpage

\begin{thebibliography}{86}%
\makeatletter
\providecommand \@ifxundefined [1]{%
 \@ifx{#1\undefined}
}%
\providecommand \@ifnum [1]{%
 \ifnum #1\expandafter \@firstoftwo
 \else \expandafter \@secondoftwo
 \fi
}%
\providecommand \@ifx [1]{%
 \ifx #1\expandafter \@firstoftwo
 \else \expandafter \@secondoftwo
 \fi
}%
\providecommand \natexlab [1]{#1}%
\providecommand \enquote  [1]{``#1''}%
\providecommand \bibnamefont  [1]{#1}%
\providecommand \bibfnamefont [1]{#1}%
\providecommand \citenamefont [1]{#1}%
\providecommand \href@noop [0]{\@secondoftwo}%
\providecommand \href [0]{\begingroup \@sanitize@url \@href}%
\providecommand \@href[1]{\@@startlink{#1}\@@href}%
\providecommand \@@href[1]{\endgroup#1\@@endlink}%
\providecommand \@sanitize@url [0]{\catcode `\\12\catcode `\$12\catcode
  `\&12\catcode `\#12\catcode `\^12\catcode `\_12\catcode `\%12\relax}%
\providecommand \@@startlink[1]{}%
\providecommand \@@endlink[0]{}%
\providecommand \url  [0]{\begingroup\@sanitize@url \@url }%
\providecommand \@url [1]{\endgroup\@href {#1}{\urlprefix }}%
\providecommand \urlprefix  [0]{URL }%
\providecommand \Eprint [0]{\href }%
\providecommand \doibase [0]{http://dx.doi.org/}%
\providecommand \selectlanguage [0]{\@gobble}%
\providecommand \bibinfo  [0]{\@secondoftwo}%
\providecommand \bibfield  [0]{\@secondoftwo}%
\providecommand \translation [1]{[#1]}%
\providecommand \BibitemOpen [0]{}%
\providecommand \bibitemStop [0]{}%
\providecommand \bibitemNoStop [0]{.\EOS\space}%
\providecommand \EOS [0]{\spacefactor3000\relax}%
\providecommand \BibitemShut  [1]{\csname bibitem#1\endcsname}%
\let\auto@bib@innerbib\@empty
\bibitem [{\citenamefont {Cordova}\ \emph {et~al.}(2022)\citenamefont
  {Cordova}, \citenamefont {Dumitrescu}, \citenamefont {Intriligator},\ and\
  \citenamefont {Shao}}]{Cordova:2022ruw}%
  \BibitemOpen
  \bibfield  {author} {\bibinfo {author} {\bibfnamefont {Clay}\ \bibnamefont
  {Cordova}}, \bibinfo {author} {\bibfnamefont {Thomas~T.}\ \bibnamefont
  {Dumitrescu}}, \bibinfo {author} {\bibfnamefont {Kenneth}\ \bibnamefont
  {Intriligator}}, \ and\ \bibinfo {author} {\bibfnamefont {Shu-Heng}\
  \bibnamefont {Shao}},\ }\bibfield  {title} {\enquote {\bibinfo {title}
  {{Snowmass White Paper: Generalized Symmetries in Quantum Field Theory and
  Beyond}},}\ }in\ \href@noop {} {\emph {\bibinfo {booktitle} {{Snowmass
  2021}}}}\ (\bibinfo {year} {2022})\ \Eprint {http://arxiv.org/abs/2205.09545}
  {arXiv:2205.09545 [hep-th]} \BibitemShut {NoStop}%
\bibitem [{\citenamefont {McGreevy}(2023)}]{McGreevy:2022oyu}%
  \BibitemOpen
  \bibfield  {author} {\bibinfo {author} {\bibfnamefont {John}\ \bibnamefont
  {McGreevy}},\ }\bibfield  {title} {\enquote {\bibinfo {title} {{Generalized
  Symmetries in Condensed Matter}},}\ }\href {\doibase
  10.1146/annurev-conmatphys-040721-021029} {\bibfield  {journal} {\bibinfo
  {journal} {Ann. Rev. Condensed Matter Phys.}\ }\textbf {\bibinfo {volume}
  {14}},\ \bibinfo {pages} {57--82} (\bibinfo {year} {2023})},\ \Eprint
  {http://arxiv.org/abs/2204.03045} {arXiv:2204.03045 [cond-mat.str-el]}
  \BibitemShut {NoStop}%
\bibitem [{\citenamefont {Brennan}\ and\ \citenamefont
  {Hong}(2023)}]{Brennan:2023mmt}%
  \BibitemOpen
  \bibfield  {author} {\bibinfo {author} {\bibfnamefont {T.~Daniel}\
  \bibnamefont {Brennan}}\ and\ \bibinfo {author} {\bibfnamefont {Sungwoo}\
  \bibnamefont {Hong}},\ }\bibfield  {title} {\enquote {\bibinfo {title}
  {{Introduction to Generalized Global Symmetries in QFT and Particle
  Physics}},}\ }\href@noop {} {\  (\bibinfo {year} {2023})},\ \Eprint
  {http://arxiv.org/abs/2306.00912} {arXiv:2306.00912 [hep-ph]} \BibitemShut
  {NoStop}%
\bibitem [{\citenamefont {Gomes}(2023)}]{Gomes:2023ahz}%
  \BibitemOpen
  \bibfield  {author} {\bibinfo {author} {\bibfnamefont {Pedro R.~S.}\
  \bibnamefont {Gomes}},\ }\bibfield  {title} {\enquote {\bibinfo {title} {{An
  introduction to higher-form symmetries}},}\ }\href {\doibase
  10.21468/SciPostPhysLectNotes.74} {\bibfield  {journal} {\bibinfo  {journal}
  {SciPost Phys. Lect. Notes}\ }\textbf {\bibinfo {volume} {74}},\ \bibinfo
  {pages} {1} (\bibinfo {year} {2023})},\ \Eprint
  {http://arxiv.org/abs/2303.01817} {arXiv:2303.01817 [hep-th]} \BibitemShut
  {NoStop}%
\bibitem [{\citenamefont {Shao}(2023)}]{Shao:2023gho}%
  \BibitemOpen
  \bibfield  {author} {\bibinfo {author} {\bibfnamefont {Shu-Heng}\
  \bibnamefont {Shao}},\ }\bibfield  {title} {\enquote {\bibinfo {title}
  {{What's Done Cannot Be Undone: TASI Lectures on Non-Invertible
  Symmetries}},}\ }\href@noop {} {\  (\bibinfo {year} {2023})},\ \Eprint
  {http://arxiv.org/abs/2308.00747} {arXiv:2308.00747 [hep-th]} \BibitemShut
  {NoStop}%
\bibitem [{\citenamefont {Schafer-Nameki}(2024)}]{Schafer-Nameki:2023jdn}%
  \BibitemOpen
  \bibfield  {author} {\bibinfo {author} {\bibfnamefont {Sakura}\ \bibnamefont
  {Schafer-Nameki}},\ }\bibfield  {title} {\enquote {\bibinfo {title} {{ICTP
  lectures on (non-)invertible generalized symmetries}},}\ }\href {\doibase
  10.1016/j.physrep.2024.01.007} {\bibfield  {journal} {\bibinfo  {journal}
  {Phys. Rept.}\ }\textbf {\bibinfo {volume} {1063}},\ \bibinfo {pages} {1--55}
  (\bibinfo {year} {2024})},\ \Eprint {http://arxiv.org/abs/2305.18296}
  {arXiv:2305.18296 [hep-th]} \BibitemShut {NoStop}%
\bibitem [{\citenamefont {Bhardwaj}\ \emph {et~al.}(2024)\citenamefont
  {Bhardwaj}, \citenamefont {Bottini}, \citenamefont {Fraser-Taliente},
  \citenamefont {Gladden}, \citenamefont {Gould}, \citenamefont {Platschorre},\
  and\ \citenamefont {Tillim}}]{Bhardwaj:2023kri}%
  \BibitemOpen
  \bibfield  {author} {\bibinfo {author} {\bibfnamefont {Lakshya}\ \bibnamefont
  {Bhardwaj}}, \bibinfo {author} {\bibfnamefont {Lea~E.}\ \bibnamefont
  {Bottini}}, \bibinfo {author} {\bibfnamefont {Ludovic}\ \bibnamefont
  {Fraser-Taliente}}, \bibinfo {author} {\bibfnamefont {Liam}\ \bibnamefont
  {Gladden}}, \bibinfo {author} {\bibfnamefont {Dewi S.~W.}\ \bibnamefont
  {Gould}}, \bibinfo {author} {\bibfnamefont {Arthur}\ \bibnamefont
  {Platschorre}}, \ and\ \bibinfo {author} {\bibfnamefont {Hannah}\
  \bibnamefont {Tillim}},\ }\bibfield  {title} {\enquote {\bibinfo {title}
  {{Lectures on generalized symmetries}},}\ }\href {\doibase
  10.1016/j.physrep.2023.11.002} {\bibfield  {journal} {\bibinfo  {journal}
  {Phys. Rept.}\ }\textbf {\bibinfo {volume} {1051}},\ \bibinfo {pages} {1--87}
  (\bibinfo {year} {2024})},\ \Eprint {http://arxiv.org/abs/2307.07547}
  {arXiv:2307.07547 [hep-th]} \BibitemShut {NoStop}%
\bibitem [{\citenamefont {Iqbal}(2024)}]{Iqbal:2024pee}%
  \BibitemOpen
  \bibfield  {author} {\bibinfo {author} {\bibfnamefont {Nabil}\ \bibnamefont
  {Iqbal}},\ }\bibfield  {title} {\enquote {\bibinfo {title} {{Jena lectures on
  generalized global symmetries: principles and applications}},}\ \ }(\bibinfo
  {year} {2024})\ \Eprint {http://arxiv.org/abs/2407.20815} {arXiv:2407.20815
  [hep-th]} \BibitemShut {NoStop}%
\bibitem [{\citenamefont {Fukushima}\ and\ \citenamefont
  {Hatsuda}(2011)}]{Fukushima:2010bq}%
  \BibitemOpen
  \bibfield  {author} {\bibinfo {author} {\bibfnamefont {Kenji}\ \bibnamefont
  {Fukushima}}\ and\ \bibinfo {author} {\bibfnamefont {Tetsuo}\ \bibnamefont
  {Hatsuda}},\ }\bibfield  {title} {\enquote {\bibinfo {title} {{The phase
  diagram of dense QCD}},}\ }\href {\doibase 10.1088/0034-4885/74/1/014001}
  {\bibfield  {journal} {\bibinfo  {journal} {Rept. Prog. Phys.}\ }\textbf
  {\bibinfo {volume} {74}},\ \bibinfo {pages} {014001} (\bibinfo {year}
  {2011})},\ \Eprint {http://arxiv.org/abs/1005.4814} {arXiv:1005.4814
  [hep-ph]} \BibitemShut {NoStop}%
\bibitem [{\citenamefont {Gour}\ \emph {et~al.}(2009)\citenamefont {Gour},
  \citenamefont {Marvian},\ and\ \citenamefont {Spekkens}}]{gour2009measuring}%
  \BibitemOpen
  \bibfield  {author} {\bibinfo {author} {\bibfnamefont {Gilad}\ \bibnamefont
  {Gour}}, \bibinfo {author} {\bibfnamefont {Iman}\ \bibnamefont {Marvian}}, \
  and\ \bibinfo {author} {\bibfnamefont {Robert~W.}\ \bibnamefont {Spekkens}},\
  }\bibfield  {title} {\enquote {\bibinfo {title} {{Measuring the quality of a
  quantum reference frame: The relative entropy of frameness}},}\ }\href
  {\doibase 10.1103/PhysRevA.80.012307} {\bibfield  {journal} {\bibinfo
  {journal} {Phys. Rev. A}\ }\textbf {\bibinfo {volume} {80}},\ \bibinfo
  {pages} {012307} (\bibinfo {year} {2009})}\BibitemShut {NoStop}%
\bibitem [{\citenamefont {Vaccaro}\ \emph {et~al.}(2008)\citenamefont
  {Vaccaro}, \citenamefont {Anselmi}, \citenamefont {Wiseman},\ and\
  \citenamefont {Jacobs}}]{vaccaro2008tradeoff}%
  \BibitemOpen
  \bibfield  {author} {\bibinfo {author} {\bibfnamefont {J.~A.}\ \bibnamefont
  {Vaccaro}}, \bibinfo {author} {\bibfnamefont {F.}~\bibnamefont {Anselmi}},
  \bibinfo {author} {\bibfnamefont {H.~M.}\ \bibnamefont {Wiseman}}, \ and\
  \bibinfo {author} {\bibfnamefont {K.}~\bibnamefont {Jacobs}},\ }\bibfield
  {title} {\enquote {\bibinfo {title} {Tradeoff between extractable mechanical
  work, accessible entanglement, and ability to act as a reference system,
  under arbitrary superselection rules},}\ }\href {\doibase
  10.1103/PhysRevA.77.032114} {\bibfield  {journal} {\bibinfo  {journal} {Phys.
  Rev. A}\ }\textbf {\bibinfo {volume} {77}},\ \bibinfo {pages} {032114}
  (\bibinfo {year} {2008})}\BibitemShut {NoStop}%
\bibitem [{\citenamefont {Ares}\ \emph {et~al.}(2023)\citenamefont {Ares},
  \citenamefont {Murciano},\ and\ \citenamefont {Calabrese}}]{Ares:2022koq}%
  \BibitemOpen
  \bibfield  {author} {\bibinfo {author} {\bibfnamefont {Filiberto}\
  \bibnamefont {Ares}}, \bibinfo {author} {\bibfnamefont {Sara}\ \bibnamefont
  {Murciano}}, \ and\ \bibinfo {author} {\bibfnamefont {Pasquale}\ \bibnamefont
  {Calabrese}},\ }\bibfield  {title} {\enquote {\bibinfo {title} {{Entanglement
  asymmetry as a probe of symmetry breaking}},}\ }\href {\doibase
  10.1038/s41467-023-37747-8} {\bibfield  {journal} {\bibinfo  {journal}
  {Nature Commun.}\ }\textbf {\bibinfo {volume} {14}},\ \bibinfo {pages} {2036}
  (\bibinfo {year} {2023})},\ \Eprint {http://arxiv.org/abs/2207.14693}
  {arXiv:2207.14693 [cond-mat.stat-mech]} \BibitemShut {NoStop}%
\bibitem [{\citenamefont {Srednicki}(1993)}]{Srednicki:1993im}%
  \BibitemOpen
  \bibfield  {author} {\bibinfo {author} {\bibfnamefont {Mark}\ \bibnamefont
  {Srednicki}},\ }\bibfield  {title} {\enquote {\bibinfo {title} {{Entropy and
  area}},}\ }\href {\doibase 10.1103/PhysRevLett.71.666} {\bibfield  {journal}
  {\bibinfo  {journal} {Phys. Rev. Lett.}\ }\textbf {\bibinfo {volume} {71}},\
  \bibinfo {pages} {666--669} (\bibinfo {year} {1993})},\ \Eprint
  {http://arxiv.org/abs/hep-th/9303048} {arXiv:hep-th/9303048} \BibitemShut
  {NoStop}%
\bibitem [{\citenamefont {Bombelli}\ \emph {et~al.}(1986)\citenamefont
  {Bombelli}, \citenamefont {Koul}, \citenamefont {Lee},\ and\ \citenamefont
  {Sorkin}}]{bombelli1986}%
  \BibitemOpen
  \bibfield  {author} {\bibinfo {author} {\bibfnamefont {Luca}\ \bibnamefont
  {Bombelli}}, \bibinfo {author} {\bibfnamefont {Rabinder~K.}\ \bibnamefont
  {Koul}}, \bibinfo {author} {\bibfnamefont {Joohan}\ \bibnamefont {Lee}}, \
  and\ \bibinfo {author} {\bibfnamefont {Rafael~D.}\ \bibnamefont {Sorkin}},\
  }\bibfield  {title} {\enquote {\bibinfo {title} {Quantum source of entropy
  for black holes},}\ }\href {\doibase 10.1103/PhysRevD.34.373} {\bibfield
  {journal} {\bibinfo  {journal} {Phys. Rev. D}\ }\textbf {\bibinfo {volume}
  {34}},\ \bibinfo {pages} {373--383} (\bibinfo {year} {1986})}\BibitemShut
  {NoStop}%
\bibitem [{\citenamefont {Calabrese}\ and\ \citenamefont
  {Cardy}(2004)}]{Calabrese:2004eu}%
  \BibitemOpen
  \bibfield  {author} {\bibinfo {author} {\bibfnamefont {Pasquale}\
  \bibnamefont {Calabrese}}\ and\ \bibinfo {author} {\bibfnamefont {John~L.}\
  \bibnamefont {Cardy}},\ }\bibfield  {title} {\enquote {\bibinfo {title}
  {{Entanglement entropy and quantum field theory}},}\ }\href {\doibase
  10.1088/1742-5468/2004/06/P06002} {\bibfield  {journal} {\bibinfo  {journal}
  {J. Stat. Mech.}\ }\textbf {\bibinfo {volume} {0406}},\ \bibinfo {pages}
  {P06002} (\bibinfo {year} {2004})},\ \Eprint
  {http://arxiv.org/abs/hep-th/0405152} {arXiv:hep-th/0405152} \BibitemShut
  {NoStop}%
\bibitem [{\citenamefont {Holzhey}\ \emph {et~al.}(1994)\citenamefont
  {Holzhey}, \citenamefont {Larsen},\ and\ \citenamefont
  {Wilczek}}]{holzhey1994}%
  \BibitemOpen
  \bibfield  {author} {\bibinfo {author} {\bibfnamefont {Christoph}\
  \bibnamefont {Holzhey}}, \bibinfo {author} {\bibfnamefont {Finn}\
  \bibnamefont {Larsen}}, \ and\ \bibinfo {author} {\bibfnamefont {Frank}\
  \bibnamefont {Wilczek}},\ }\bibfield  {title} {\enquote {\bibinfo {title}
  {Geometric and renormalized entropy in conformal field theory},}\ }\href
  {\doibase https://doi.org/10.1016/0550-3213(94)90402-2} {\bibfield  {journal}
  {\bibinfo  {journal} {Nuclear Physics B}\ }\textbf {\bibinfo {volume}
  {424}},\ \bibinfo {pages} {443--467} (\bibinfo {year} {1994})}\BibitemShut
  {NoStop}%
\bibitem [{\citenamefont {Ryu}\ and\ \citenamefont
  {Takayanagi}(2006)}]{Ryu:2006bv}%
  \BibitemOpen
  \bibfield  {author} {\bibinfo {author} {\bibfnamefont {Shinsei}\ \bibnamefont
  {Ryu}}\ and\ \bibinfo {author} {\bibfnamefont {Tadashi}\ \bibnamefont
  {Takayanagi}},\ }\bibfield  {title} {\enquote {\bibinfo {title} {{Holographic
  derivation of entanglement entropy from AdS/CFT}},}\ }\href {\doibase
  10.1103/PhysRevLett.96.181602} {\bibfield  {journal} {\bibinfo  {journal}
  {Phys. Rev. Lett.}\ }\textbf {\bibinfo {volume} {96}},\ \bibinfo {pages}
  {181602} (\bibinfo {year} {2006})},\ \Eprint
  {http://arxiv.org/abs/hep-th/0603001} {arXiv:hep-th/0603001} \BibitemShut
  {NoStop}%
\bibitem [{\citenamefont {Cervera-Lierta}\ \emph {et~al.}(2017)\citenamefont
  {Cervera-Lierta}, \citenamefont {Latorre}, \citenamefont {Rojo},\ and\
  \citenamefont {Rottoli}}]{Cervera-Lierta:2017tdt}%
  \BibitemOpen
  \bibfield  {author} {\bibinfo {author} {\bibfnamefont {Alba}\ \bibnamefont
  {Cervera-Lierta}}, \bibinfo {author} {\bibfnamefont {Jos{\'e}~I.}\
  \bibnamefont {Latorre}}, \bibinfo {author} {\bibfnamefont {Juan}\
  \bibnamefont {Rojo}}, \ and\ \bibinfo {author} {\bibfnamefont {Luca}\
  \bibnamefont {Rottoli}},\ }\bibfield  {title} {\enquote {\bibinfo {title}
  {{Maximal Entanglement in High Energy Physics}},}\ }\href {\doibase
  10.21468/SciPostPhys.3.5.036} {\bibfield  {journal} {\bibinfo  {journal}
  {SciPost Phys.}\ }\textbf {\bibinfo {volume} {3}},\ \bibinfo {pages} {036}
  (\bibinfo {year} {2017})},\ \Eprint {http://arxiv.org/abs/1703.02989}
  {arXiv:1703.02989 [hep-th]} \BibitemShut {NoStop}%
\bibitem [{\citenamefont {Kharzeev}\ and\ \citenamefont
  {Levin}(2017)}]{Kharzeev:2017qzs}%
  \BibitemOpen
  \bibfield  {author} {\bibinfo {author} {\bibfnamefont {Dmitri~E.}\
  \bibnamefont {Kharzeev}}\ and\ \bibinfo {author} {\bibfnamefont {Eugene~M.}\
  \bibnamefont {Levin}},\ }\bibfield  {title} {\enquote {\bibinfo {title}
  {{Deep inelastic scattering as a probe of entanglement}},}\ }\href {\doibase
  10.1103/PhysRevD.95.114008} {\bibfield  {journal} {\bibinfo  {journal} {Phys.
  Rev. D}\ }\textbf {\bibinfo {volume} {95}},\ \bibinfo {pages} {114008}
  (\bibinfo {year} {2017})},\ \Eprint {http://arxiv.org/abs/1702.03489}
  {arXiv:1702.03489 [hep-ph]} \BibitemShut {NoStop}%
\bibitem [{\citenamefont {Afik}\ and\ \citenamefont
  {de~Nova}(2021)}]{Afik:2020onf}%
  \BibitemOpen
  \bibfield  {author} {\bibinfo {author} {\bibfnamefont {Yoav}\ \bibnamefont
  {Afik}}\ and\ \bibinfo {author} {\bibfnamefont {Juan Ram{\'o}n~Mu{\~n}oz}\
  \bibnamefont {de~Nova}},\ }\bibfield  {title} {\enquote {\bibinfo {title}
  {{Entanglement and quantum tomography with top quarks at the LHC}},}\ }\href
  {\doibase 10.1140/epjp/s13360-021-01902-1} {\bibfield  {journal} {\bibinfo
  {journal} {Eur. Phys. J. Plus}\ }\textbf {\bibinfo {volume} {136}},\ \bibinfo
  {pages} {907} (\bibinfo {year} {2021})},\ \Eprint
  {http://arxiv.org/abs/2003.02280} {arXiv:2003.02280 [quant-ph]} \BibitemShut
  {NoStop}%
\bibitem [{\citenamefont {Florio}\ and\ \citenamefont
  {Kharzeev}(2021)}]{Florio:2021xvj}%
  \BibitemOpen
  \bibfield  {author} {\bibinfo {author} {\bibfnamefont {Adrien}\ \bibnamefont
  {Florio}}\ and\ \bibinfo {author} {\bibfnamefont {Dmitri~E.}\ \bibnamefont
  {Kharzeev}},\ }\bibfield  {title} {\enquote {\bibinfo {title} {{Gibbs entropy
  from entanglement in electric quenches}},}\ }\href {\doibase
  10.1103/PhysRevD.104.056021} {\bibfield  {journal} {\bibinfo  {journal}
  {Phys. Rev. D}\ }\textbf {\bibinfo {volume} {104}},\ \bibinfo {pages}
  {056021} (\bibinfo {year} {2021})},\ \Eprint
  {http://arxiv.org/abs/2106.00838} {arXiv:2106.00838 [hep-th]} \BibitemShut
  {NoStop}%
\bibitem [{\citenamefont {Kharzeev}\ and\ \citenamefont
  {Levin}(2021)}]{Kharzeev:2021yyf}%
  \BibitemOpen
  \bibfield  {author} {\bibinfo {author} {\bibfnamefont {Dmitri~E.}\
  \bibnamefont {Kharzeev}}\ and\ \bibinfo {author} {\bibfnamefont {Eugene}\
  \bibnamefont {Levin}},\ }\bibfield  {title} {\enquote {\bibinfo {title}
  {{Deep inelastic scattering as a probe of entanglement: Confronting
  experimental data}},}\ }\href {\doibase 10.1103/PhysRevD.104.L031503}
  {\bibfield  {journal} {\bibinfo  {journal} {Phys. Rev. D}\ }\textbf {\bibinfo
  {volume} {104}},\ \bibinfo {pages} {L031503} (\bibinfo {year} {2021})},\
  \Eprint {http://arxiv.org/abs/2102.09773} {arXiv:2102.09773 [hep-ph]}
  \BibitemShut {NoStop}%
\bibitem [{\citenamefont {Gong}\ \emph {et~al.}(2022)\citenamefont {Gong},
  \citenamefont {Parida}, \citenamefont {Tu},\ and\ \citenamefont
  {Venugopalan}}]{Gong:2021bcp}%
  \BibitemOpen
  \bibfield  {author} {\bibinfo {author} {\bibfnamefont {Wenjie}\ \bibnamefont
  {Gong}}, \bibinfo {author} {\bibfnamefont {Ganesh}\ \bibnamefont {Parida}},
  \bibinfo {author} {\bibfnamefont {Zhoudunming}\ \bibnamefont {Tu}}, \ and\
  \bibinfo {author} {\bibfnamefont {Raju}\ \bibnamefont {Venugopalan}},\
  }\bibfield  {title} {\enquote {\bibinfo {title} {{Measurement of Bell-type
  inequalities and quantum entanglement from {\ensuremath{\Lambda}}-hyperon
  spin correlations at high energy colliders}},}\ }\href {\doibase
  10.1103/PhysRevD.106.L031501} {\bibfield  {journal} {\bibinfo  {journal}
  {Phys. Rev. D}\ }\textbf {\bibinfo {volume} {106}},\ \bibinfo {pages}
  {L031501} (\bibinfo {year} {2022})},\ \Eprint
  {http://arxiv.org/abs/2107.13007} {arXiv:2107.13007 [hep-ph]} \BibitemShut
  {NoStop}%
\bibitem [{\citenamefont {de~Jong}\ \emph {et~al.}(2022)\citenamefont
  {de~Jong}, \citenamefont {Lee}, \citenamefont {Mulligan}, \citenamefont
  {P{\l}osko{\'n}}, \citenamefont {Ringer},\ and\ \citenamefont
  {Yao}}]{deJong:2021wsd}%
  \BibitemOpen
  \bibfield  {author} {\bibinfo {author} {\bibfnamefont {Wibe~A.}\ \bibnamefont
  {de~Jong}}, \bibinfo {author} {\bibfnamefont {Kyle}\ \bibnamefont {Lee}},
  \bibinfo {author} {\bibfnamefont {James}\ \bibnamefont {Mulligan}}, \bibinfo
  {author} {\bibfnamefont {Mateusz}\ \bibnamefont {P{\l}osko{\'n}}}, \bibinfo
  {author} {\bibfnamefont {Felix}\ \bibnamefont {Ringer}}, \ and\ \bibinfo
  {author} {\bibfnamefont {Xiaojun}\ \bibnamefont {Yao}},\ }\bibfield  {title}
  {\enquote {\bibinfo {title} {{Quantum simulation of nonequilibrium dynamics
  and thermalization in the Schwinger model}},}\ }\href {\doibase
  10.1103/PhysRevD.106.054508} {\bibfield  {journal} {\bibinfo  {journal}
  {Phys. Rev. D}\ }\textbf {\bibinfo {volume} {106}},\ \bibinfo {pages}
  {054508} (\bibinfo {year} {2022})},\ \Eprint
  {http://arxiv.org/abs/2106.08394} {arXiv:2106.08394 [quant-ph]} \BibitemShut
  {NoStop}%
\bibitem [{\citenamefont {Dunne}\ \emph {et~al.}(2023)\citenamefont {Dunne},
  \citenamefont {Florio},\ and\ \citenamefont {Kharzeev}}]{Dunne:2022zlx}%
  \BibitemOpen
  \bibfield  {author} {\bibinfo {author} {\bibfnamefont {Gerald~V.}\
  \bibnamefont {Dunne}}, \bibinfo {author} {\bibfnamefont {Adrien}\
  \bibnamefont {Florio}}, \ and\ \bibinfo {author} {\bibfnamefont {Dmitri~E.}\
  \bibnamefont {Kharzeev}},\ }\bibfield  {title} {\enquote {\bibinfo {title}
  {{Entropy suppression through quantum interference in electric pulses}},}\
  }\href {\doibase 10.1103/PhysRevD.108.L031901} {\bibfield  {journal}
  {\bibinfo  {journal} {Phys. Rev. D}\ }\textbf {\bibinfo {volume} {108}},\
  \bibinfo {pages} {L031901} (\bibinfo {year} {2023})},\ \Eprint
  {http://arxiv.org/abs/2211.13347} {arXiv:2211.13347 [hep-ph]} \BibitemShut
  {NoStop}%
\bibitem [{\citenamefont {Florio}\ \emph {et~al.}(2023)\citenamefont {Florio},
  \citenamefont {Frenklakh}, \citenamefont {Ikeda}, \citenamefont {Kharzeev},
  \citenamefont {Korepin}, \citenamefont {Shi},\ and\ \citenamefont
  {Yu}}]{Florio:2023dke}%
  \BibitemOpen
  \bibfield  {author} {\bibinfo {author} {\bibfnamefont {Adrien}\ \bibnamefont
  {Florio}}, \bibinfo {author} {\bibfnamefont {David}\ \bibnamefont
  {Frenklakh}}, \bibinfo {author} {\bibfnamefont {Kazuki}\ \bibnamefont
  {Ikeda}}, \bibinfo {author} {\bibfnamefont {Dmitri}\ \bibnamefont
  {Kharzeev}}, \bibinfo {author} {\bibfnamefont {Vladimir}\ \bibnamefont
  {Korepin}}, \bibinfo {author} {\bibfnamefont {Shuzhe}\ \bibnamefont {Shi}}, \
  and\ \bibinfo {author} {\bibfnamefont {Kwangmin}\ \bibnamefont {Yu}},\
  }\bibfield  {title} {\enquote {\bibinfo {title} {{Real-Time Nonperturbative
  Dynamics of Jet Production in Schwinger Model: Quantum Entanglement and
  Vacuum Modification}},}\ }\href {\doibase 10.1103/PhysRevLett.131.021902}
  {\bibfield  {journal} {\bibinfo  {journal} {Phys. Rev. Lett.}\ }\textbf
  {\bibinfo {volume} {131}},\ \bibinfo {pages} {021902} (\bibinfo {year}
  {2023})},\ \Eprint {http://arxiv.org/abs/2301.11991} {arXiv:2301.11991
  [hep-ph]} \BibitemShut {NoStop}%
\bibitem [{\citenamefont {Belyansky}\ \emph {et~al.}(2024)\citenamefont
  {Belyansky}, \citenamefont {Whitsitt}, \citenamefont {Mueller}, \citenamefont
  {Fahimniya}, \citenamefont {Bennewitz}, \citenamefont {Davoudi},\ and\
  \citenamefont {Gorshkov}}]{Belyansky:2023rgh}%
  \BibitemOpen
  \bibfield  {author} {\bibinfo {author} {\bibfnamefont {Ron}\ \bibnamefont
  {Belyansky}}, \bibinfo {author} {\bibfnamefont {Seth}\ \bibnamefont
  {Whitsitt}}, \bibinfo {author} {\bibfnamefont {Niklas}\ \bibnamefont
  {Mueller}}, \bibinfo {author} {\bibfnamefont {Ali}\ \bibnamefont
  {Fahimniya}}, \bibinfo {author} {\bibfnamefont {Elizabeth~R.}\ \bibnamefont
  {Bennewitz}}, \bibinfo {author} {\bibfnamefont {Zohreh}\ \bibnamefont
  {Davoudi}}, \ and\ \bibinfo {author} {\bibfnamefont {Alexey~V.}\ \bibnamefont
  {Gorshkov}},\ }\bibfield  {title} {\enquote {\bibinfo {title} {{High-Energy
  Collision of Quarks and Mesons in the Schwinger Model: From Tensor Networks
  to Circuit QED}},}\ }\href {\doibase 10.1103/PhysRevLett.132.091903}
  {\bibfield  {journal} {\bibinfo  {journal} {Phys. Rev. Lett.}\ }\textbf
  {\bibinfo {volume} {132}},\ \bibinfo {pages} {091903} (\bibinfo {year}
  {2024})},\ \Eprint {http://arxiv.org/abs/2307.02522} {arXiv:2307.02522
  [quant-ph]} \BibitemShut {NoStop}%
\bibitem [{\citenamefont {Barata}\ \emph {et~al.}(2024)\citenamefont {Barata},
  \citenamefont {Gong},\ and\ \citenamefont {Venugopalan}}]{Barata:2023jgd}%
  \BibitemOpen
  \bibfield  {author} {\bibinfo {author} {\bibfnamefont {Jo{\~a}o}\
  \bibnamefont {Barata}}, \bibinfo {author} {\bibfnamefont {Wenjie}\
  \bibnamefont {Gong}}, \ and\ \bibinfo {author} {\bibfnamefont {Raju}\
  \bibnamefont {Venugopalan}},\ }\bibfield  {title} {\enquote {\bibinfo {title}
  {{Realtime dynamics of hyperon spin correlations from string fragmentation in
  a deformed four-flavor Schwinger model}},}\ }\href {\doibase
  10.1103/PhysRevD.109.116003} {\bibfield  {journal} {\bibinfo  {journal}
  {Phys. Rev. D}\ }\textbf {\bibinfo {volume} {109}},\ \bibinfo {pages}
  {116003} (\bibinfo {year} {2024})},\ \Eprint
  {http://arxiv.org/abs/2308.13596} {arXiv:2308.13596 [hep-ph]} \BibitemShut
  {NoStop}%
\bibitem [{\citenamefont {Grieninger}\ \emph
  {et~al.}(2023{\natexlab{a}})\citenamefont {Grieninger}, \citenamefont
  {Kharzeev},\ and\ \citenamefont {Zahed}}]{Grieninger:2023ehb}%
  \BibitemOpen
  \bibfield  {author} {\bibinfo {author} {\bibfnamefont {Sebastian}\
  \bibnamefont {Grieninger}}, \bibinfo {author} {\bibfnamefont {Dmitri~E.}\
  \bibnamefont {Kharzeev}}, \ and\ \bibinfo {author} {\bibfnamefont {Ismail}\
  \bibnamefont {Zahed}},\ }\bibfield  {title} {\enquote {\bibinfo {title}
  {{Entanglement in a holographic Schwinger pair with confinement}},}\ }\href
  {\doibase 10.1103/PhysRevD.108.086030} {\bibfield  {journal} {\bibinfo
  {journal} {Phys. Rev. D}\ }\textbf {\bibinfo {volume} {108}},\ \bibinfo
  {pages} {086030} (\bibinfo {year} {2023}{\natexlab{a}})},\ \Eprint
  {http://arxiv.org/abs/2305.07121} {arXiv:2305.07121 [hep-th]} \BibitemShut
  {NoStop}%
\bibitem [{\citenamefont {Grieninger}\ \emph
  {et~al.}(2023{\natexlab{b}})\citenamefont {Grieninger}, \citenamefont
  {Kharzeev},\ and\ \citenamefont {Zahed}}]{Grieninger:2023pyb}%
  \BibitemOpen
  \bibfield  {author} {\bibinfo {author} {\bibfnamefont {Sebastian}\
  \bibnamefont {Grieninger}}, \bibinfo {author} {\bibfnamefont {Dmitri~E.}\
  \bibnamefont {Kharzeev}}, \ and\ \bibinfo {author} {\bibfnamefont {Ismail}\
  \bibnamefont {Zahed}},\ }\bibfield  {title} {\enquote {\bibinfo {title}
  {{Entanglement entropy in a time-dependent holographic Schwinger pair
  creation}},}\ }\href {\doibase 10.1103/PhysRevD.108.126014} {\bibfield
  {journal} {\bibinfo  {journal} {Phys. Rev. D}\ }\textbf {\bibinfo {volume}
  {108}},\ \bibinfo {pages} {126014} (\bibinfo {year} {2023}{\natexlab{b}})},\
  \Eprint {http://arxiv.org/abs/2310.12042} {arXiv:2310.12042 [hep-th]}
  \BibitemShut {NoStop}%
\bibitem [{\citenamefont {Lee}\ \emph {et~al.}(2023)\citenamefont {Lee},
  \citenamefont {Mulligan}, \citenamefont {Ringer},\ and\ \citenamefont
  {Yao}}]{Lee:2023urk}%
  \BibitemOpen
  \bibfield  {author} {\bibinfo {author} {\bibfnamefont {Kyle}\ \bibnamefont
  {Lee}}, \bibinfo {author} {\bibfnamefont {James}\ \bibnamefont {Mulligan}},
  \bibinfo {author} {\bibfnamefont {Felix}\ \bibnamefont {Ringer}}, \ and\
  \bibinfo {author} {\bibfnamefont {Xiaojun}\ \bibnamefont {Yao}},\ }\bibfield
  {title} {\enquote {\bibinfo {title} {{Liouvillian dynamics of the open
  Schwinger model: String breaking and kinetic dissipation in a thermal
  medium}},}\ }\href {\doibase 10.1103/PhysRevD.108.094518} {\bibfield
  {journal} {\bibinfo  {journal} {Phys. Rev. D}\ }\textbf {\bibinfo {volume}
  {108}},\ \bibinfo {pages} {094518} (\bibinfo {year} {2023})},\ \Eprint
  {http://arxiv.org/abs/2308.03878} {arXiv:2308.03878 [quant-ph]} \BibitemShut
  {NoStop}%
\bibitem [{\citenamefont {Kirchner}\ \emph {et~al.}(2024)\citenamefont
  {Kirchner}, \citenamefont {Elkamhawy},\ and\ \citenamefont
  {Hammer}}]{Kirchner:2023dvg}%
  \BibitemOpen
  \bibfield  {author} {\bibinfo {author} {\bibfnamefont {Tanja}\ \bibnamefont
  {Kirchner}}, \bibinfo {author} {\bibfnamefont {Wael}\ \bibnamefont
  {Elkamhawy}}, \ and\ \bibinfo {author} {\bibfnamefont {Hans-Werner}\
  \bibnamefont {Hammer}},\ }\bibfield  {title} {\enquote {\bibinfo {title}
  {{Entanglement in Few-Nucleon Scattering Events}},}\ }\href {\doibase
  10.1007/s00601-024-01897-2} {\bibfield  {journal} {\bibinfo  {journal} {Few
  Body Syst.}\ }\textbf {\bibinfo {volume} {65}},\ \bibinfo {pages} {29}
  (\bibinfo {year} {2024})},\ \Eprint {http://arxiv.org/abs/2312.14484}
  {arXiv:2312.14484 [nucl-th]} \BibitemShut {NoStop}%
\bibitem [{\citenamefont {Robin}\ and\ \citenamefont
  {Savage}(2025)}]{Robin:2025ymq}%
  \BibitemOpen
  \bibfield  {author} {\bibinfo {author} {\bibfnamefont {C.~E.~P.}\
  \bibnamefont {Robin}}\ and\ \bibinfo {author} {\bibfnamefont {M.~J.}\
  \bibnamefont {Savage}},\ }\bibfield  {title} {\enquote {\bibinfo {title}
  {{Anti-Flatness and Non-Local Magic in Two-Particle Scattering Processes}},}\
  }\href@noop {} {\  (\bibinfo {year} {2025})},\ \Eprint
  {http://arxiv.org/abs/2510.23426} {arXiv:2510.23426 [quant-ph]} \BibitemShut
  {NoStop}%
\bibitem [{\citenamefont {Beane}\ \emph {et~al.}(2019)\citenamefont {Beane},
  \citenamefont {Kaplan}, \citenamefont {Klco},\ and\ \citenamefont
  {Savage}}]{Beane:2018oxh}%
  \BibitemOpen
  \bibfield  {author} {\bibinfo {author} {\bibfnamefont {Silas~R.}\
  \bibnamefont {Beane}}, \bibinfo {author} {\bibfnamefont {David~B.}\
  \bibnamefont {Kaplan}}, \bibinfo {author} {\bibfnamefont {Natalie}\
  \bibnamefont {Klco}}, \ and\ \bibinfo {author} {\bibfnamefont {Martin~J.}\
  \bibnamefont {Savage}},\ }\bibfield  {title} {\enquote {\bibinfo {title}
  {{Entanglement Suppression and Emergent Symmetries of Strong
  Interactions}},}\ }\href {\doibase 10.1103/PhysRevLett.122.102001} {\bibfield
   {journal} {\bibinfo  {journal} {Phys. Rev. Lett.}\ }\textbf {\bibinfo
  {volume} {122}},\ \bibinfo {pages} {102001} (\bibinfo {year} {2019})},\
  \Eprint {http://arxiv.org/abs/1812.03138} {arXiv:1812.03138 [nucl-th]}
  \BibitemShut {NoStop}%
\bibitem [{\citenamefont {Klco}\ and\ \citenamefont
  {Savage}(2021{\natexlab{a}})}]{Klco:2020rga}%
  \BibitemOpen
  \bibfield  {author} {\bibinfo {author} {\bibfnamefont {Natalie}\ \bibnamefont
  {Klco}}\ and\ \bibinfo {author} {\bibfnamefont {Martin~J.}\ \bibnamefont
  {Savage}},\ }\bibfield  {title} {\enquote {\bibinfo {title} {{Geometric
  quantum information structure in quantum fields and their lattice
  simulation}},}\ }\href {\doibase 10.1103/PhysRevD.103.065007} {\bibfield
  {journal} {\bibinfo  {journal} {Phys. Rev. D}\ }\textbf {\bibinfo {volume}
  {103}},\ \bibinfo {pages} {065007} (\bibinfo {year} {2021}{\natexlab{a}})},\
  \Eprint {http://arxiv.org/abs/2008.03647} {arXiv:2008.03647 [quant-ph]}
  \BibitemShut {NoStop}%
\bibitem [{\citenamefont {Klco}\ and\ \citenamefont
  {Savage}(2021{\natexlab{b}})}]{Klco:2021biu}%
  \BibitemOpen
  \bibfield  {author} {\bibinfo {author} {\bibfnamefont {Natalie}\ \bibnamefont
  {Klco}}\ and\ \bibinfo {author} {\bibfnamefont {Martin~J.}\ \bibnamefont
  {Savage}},\ }\bibfield  {title} {\enquote {\bibinfo {title} {{Entanglement
  Spheres and a UV-IR Connection in Effective Field Theories}},}\ }\href
  {\doibase 10.1103/PhysRevLett.127.211602} {\bibfield  {journal} {\bibinfo
  {journal} {Phys. Rev. Lett.}\ }\textbf {\bibinfo {volume} {127}},\ \bibinfo
  {pages} {211602} (\bibinfo {year} {2021}{\natexlab{b}})},\ \Eprint
  {http://arxiv.org/abs/2103.14999} {arXiv:2103.14999 [hep-th]} \BibitemShut
  {NoStop}%
\bibitem [{\citenamefont {Klco}\ \emph {et~al.}(2023)\citenamefont {Klco},
  \citenamefont {Beck},\ and\ \citenamefont {Savage}}]{Klco:2021cxq}%
  \BibitemOpen
  \bibfield  {author} {\bibinfo {author} {\bibfnamefont {Natalie}\ \bibnamefont
  {Klco}}, \bibinfo {author} {\bibfnamefont {D.~H.}\ \bibnamefont {Beck}}, \
  and\ \bibinfo {author} {\bibfnamefont {Martin~J.}\ \bibnamefont {Savage}},\
  }\bibfield  {title} {\enquote {\bibinfo {title} {{Entanglement structures in
  quantum field theories: Negativity cores and bound entanglement in the
  vacuum}},}\ }\href {\doibase 10.1103/PhysRevA.107.012415} {\bibfield
  {journal} {\bibinfo  {journal} {Phys. Rev. A}\ }\textbf {\bibinfo {volume}
  {107}},\ \bibinfo {pages} {012415} (\bibinfo {year} {2023})},\ \Eprint
  {http://arxiv.org/abs/2110.10736} {arXiv:2110.10736 [quant-ph]} \BibitemShut
  {NoStop}%
\bibitem [{\citenamefont {Kharzeev}\ and\ \citenamefont
  {Tuchin}(2005)}]{Kharzeev:2005iz}%
  \BibitemOpen
  \bibfield  {author} {\bibinfo {author} {\bibfnamefont {Dmitri}\ \bibnamefont
  {Kharzeev}}\ and\ \bibinfo {author} {\bibfnamefont {Kirill}\ \bibnamefont
  {Tuchin}},\ }\bibfield  {title} {\enquote {\bibinfo {title} {{From color
  glass condensate to quark gluon plasma through the event horizon}},}\ }\href
  {\doibase 10.1016/j.nuclphysa.2005.03.001} {\bibfield  {journal} {\bibinfo
  {journal} {Nucl. Phys. A}\ }\textbf {\bibinfo {volume} {753}},\ \bibinfo
  {pages} {316--334} (\bibinfo {year} {2005})},\ \Eprint
  {http://arxiv.org/abs/hep-ph/0501234} {arXiv:hep-ph/0501234} \BibitemShut
  {NoStop}%
\bibitem [{\citenamefont {Berges}\ \emph {et~al.}(2018)\citenamefont {Berges},
  \citenamefont {Floerchinger},\ and\ \citenamefont
  {Venugopalan}}]{Berges:2017hne}%
  \BibitemOpen
  \bibfield  {author} {\bibinfo {author} {\bibfnamefont {J{\"u}rgen}\
  \bibnamefont {Berges}}, \bibinfo {author} {\bibfnamefont {Stefan}\
  \bibnamefont {Floerchinger}}, \ and\ \bibinfo {author} {\bibfnamefont {Raju}\
  \bibnamefont {Venugopalan}},\ }\bibfield  {title} {\enquote {\bibinfo {title}
  {{Dynamics of entanglement in expanding quantum fields}},}\ }\href {\doibase
  10.1007/JHEP04(2018)145} {\bibfield  {journal} {\bibinfo  {journal} {JHEP}\
  }\textbf {\bibinfo {volume} {04}},\ \bibinfo {pages} {145} (\bibinfo {year}
  {2018})},\ \Eprint {http://arxiv.org/abs/1712.09362} {arXiv:1712.09362
  [hep-th]} \BibitemShut {NoStop}%
\bibitem [{\citenamefont {Zhou}\ \emph {et~al.}(2022)\citenamefont {Zhou},
  \citenamefont {Su}, \citenamefont {Halimeh}, \citenamefont {Ott},
  \citenamefont {Sun}, \citenamefont {Hauke}, \citenamefont {Yang},
  \citenamefont {Yuan}, \citenamefont {Berges},\ and\ \citenamefont
  {Pan}}]{Zhou:2021kdl}%
  \BibitemOpen
  \bibfield  {author} {\bibinfo {author} {\bibfnamefont {Zhao-Yu}\ \bibnamefont
  {Zhou}}, \bibinfo {author} {\bibfnamefont {Guo-Xian}\ \bibnamefont {Su}},
  \bibinfo {author} {\bibfnamefont {Jad~C.}\ \bibnamefont {Halimeh}}, \bibinfo
  {author} {\bibfnamefont {Robert}\ \bibnamefont {Ott}}, \bibinfo {author}
  {\bibfnamefont {Hui}\ \bibnamefont {Sun}}, \bibinfo {author} {\bibfnamefont
  {Philipp}\ \bibnamefont {Hauke}}, \bibinfo {author} {\bibfnamefont {Bing}\
  \bibnamefont {Yang}}, \bibinfo {author} {\bibfnamefont {Zhen-Sheng}\
  \bibnamefont {Yuan}}, \bibinfo {author} {\bibfnamefont {J{\"u}rgen}\
  \bibnamefont {Berges}}, \ and\ \bibinfo {author} {\bibfnamefont {Jian-Wei}\
  \bibnamefont {Pan}},\ }\bibfield  {title} {\enquote {\bibinfo {title}
  {{Thermalization dynamics of a gauge theory on a quantum simulator}},}\
  }\href {\doibase 10.1126/science.abl6277} {\bibfield  {journal} {\bibinfo
  {journal} {Science}\ }\textbf {\bibinfo {volume} {377}},\ \bibinfo {pages}
  {abl6277} (\bibinfo {year} {2022})},\ \Eprint
  {http://arxiv.org/abs/2107.13563} {arXiv:2107.13563 [cond-mat.quant-gas]}
  \BibitemShut {NoStop}%
\bibitem [{\citenamefont {Mueller}\ \emph {et~al.}(2022)\citenamefont
  {Mueller}, \citenamefont {Zache},\ and\ \citenamefont
  {Ott}}]{Mueller:2021gxd}%
  \BibitemOpen
  \bibfield  {author} {\bibinfo {author} {\bibfnamefont {Niklas}\ \bibnamefont
  {Mueller}}, \bibinfo {author} {\bibfnamefont {Torsten~V.}\ \bibnamefont
  {Zache}}, \ and\ \bibinfo {author} {\bibfnamefont {Robert}\ \bibnamefont
  {Ott}},\ }\bibfield  {title} {\enquote {\bibinfo {title} {{Thermalization of
  Gauge Theories from their Entanglement Spectrum}},}\ }\href {\doibase
  10.1103/PhysRevLett.129.011601} {\bibfield  {journal} {\bibinfo  {journal}
  {Phys. Rev. Lett.}\ }\textbf {\bibinfo {volume} {129}},\ \bibinfo {pages}
  {011601} (\bibinfo {year} {2022})},\ \Eprint
  {http://arxiv.org/abs/2107.11416} {arXiv:2107.11416 [quant-ph]} \BibitemShut
  {NoStop}%
\bibitem [{\citenamefont {Ebner}\ \emph {et~al.}(2024)\citenamefont {Ebner},
  \citenamefont {M{\"u}ller}, \citenamefont {Sch{\"a}fer}, \citenamefont
  {Seidl},\ and\ \citenamefont {Yao}}]{Ebner:2023ixq}%
  \BibitemOpen
  \bibfield  {author} {\bibinfo {author} {\bibfnamefont {Lukas}\ \bibnamefont
  {Ebner}}, \bibinfo {author} {\bibfnamefont {Berndt}\ \bibnamefont
  {M{\"u}ller}}, \bibinfo {author} {\bibfnamefont {Andreas}\ \bibnamefont
  {Sch{\"a}fer}}, \bibinfo {author} {\bibfnamefont {Clemens}\ \bibnamefont
  {Seidl}}, \ and\ \bibinfo {author} {\bibfnamefont {Xiaojun}\ \bibnamefont
  {Yao}},\ }\bibfield  {title} {\enquote {\bibinfo {title} {{Eigenstate
  thermalization in (2+1)-dimensional SU(2) lattice gauge theory}},}\ }\href
  {\doibase 10.1103/PhysRevD.109.014504} {\bibfield  {journal} {\bibinfo
  {journal} {Phys. Rev. D}\ }\textbf {\bibinfo {volume} {109}},\ \bibinfo
  {pages} {014504} (\bibinfo {year} {2024})},\ \Eprint
  {http://arxiv.org/abs/2308.16202} {arXiv:2308.16202 [hep-lat]} \BibitemShut
  {NoStop}%
\bibitem [{\citenamefont {Yao}(2023)}]{Yao:2023pht}%
  \BibitemOpen
  \bibfield  {author} {\bibinfo {author} {\bibfnamefont {Xiaojun}\ \bibnamefont
  {Yao}},\ }\bibfield  {title} {\enquote {\bibinfo {title} {{SU(2) gauge theory
  in 2+1 dimensions on a plaquette chain obeys the eigenstate thermalization
  hypothesis}},}\ }\href {\doibase 10.1103/PhysRevD.108.L031504} {\bibfield
  {journal} {\bibinfo  {journal} {Phys. Rev. D}\ }\textbf {\bibinfo {volume}
  {108}},\ \bibinfo {pages} {L031504} (\bibinfo {year} {2023})},\ \Eprint
  {http://arxiv.org/abs/2303.14264} {arXiv:2303.14264 [hep-lat]} \BibitemShut
  {NoStop}%
\bibitem [{\citenamefont {Grieninger}\ \emph {et~al.}(2024)\citenamefont
  {Grieninger}, \citenamefont {Ikeda}, \citenamefont {Kharzeev},\ and\
  \citenamefont {Zahed}}]{Grieninger:2023ufa}%
  \BibitemOpen
  \bibfield  {author} {\bibinfo {author} {\bibfnamefont {Sebastian}\
  \bibnamefont {Grieninger}}, \bibinfo {author} {\bibfnamefont {Kazuki}\
  \bibnamefont {Ikeda}}, \bibinfo {author} {\bibfnamefont {Dmitri~E.}\
  \bibnamefont {Kharzeev}}, \ and\ \bibinfo {author} {\bibfnamefont {Ismail}\
  \bibnamefont {Zahed}},\ }\bibfield  {title} {\enquote {\bibinfo {title}
  {{Entanglement in massive Schwinger model at finite temperature and
  density}},}\ }\href {\doibase 10.1103/PhysRevD.109.016023} {\bibfield
  {journal} {\bibinfo  {journal} {Phys. Rev. D}\ }\textbf {\bibinfo {volume}
  {109}},\ \bibinfo {pages} {016023} (\bibinfo {year} {2024})},\ \Eprint
  {http://arxiv.org/abs/2312.03172} {arXiv:2312.03172 [hep-th]} \BibitemShut
  {NoStop}%
\bibitem [{\citenamefont {Ciavarella}\ \emph {et~al.}(2025)\citenamefont
  {Ciavarella}, \citenamefont {Bauer},\ and\ \citenamefont
  {Halimeh}}]{Ciavarella:2025zqf}%
  \BibitemOpen
  \bibfield  {author} {\bibinfo {author} {\bibfnamefont {Anthony~N.}\
  \bibnamefont {Ciavarella}}, \bibinfo {author} {\bibfnamefont {Christian~W.}\
  \bibnamefont {Bauer}}, \ and\ \bibinfo {author} {\bibfnamefont {Jad~C.}\
  \bibnamefont {Halimeh}},\ }\bibfield  {title} {\enquote {\bibinfo {title}
  {{Generic Hilbert Space Fragmentation in Kogut--Susskind Lattice Gauge
  Theories}},}\ }\href@noop {} {\  (\bibinfo {year} {2025})},\ \Eprint
  {http://arxiv.org/abs/2502.03533} {arXiv:2502.03533 [quant-ph]} \BibitemShut
  {NoStop}%
\bibitem [{\citenamefont {Turro}\ and\ \citenamefont
  {Yao}(2025)}]{Turro:2025sec}%
  \BibitemOpen
  \bibfield  {author} {\bibinfo {author} {\bibfnamefont {Francesco}\
  \bibnamefont {Turro}}\ and\ \bibinfo {author} {\bibfnamefont {Xiaojun}\
  \bibnamefont {Yao}},\ }\bibfield  {title} {\enquote {\bibinfo {title}
  {{Emergent hydrodynamic mode on SU(2) plaquette chains and quantum
  simulation}},}\ }\href {\doibase 10.1103/PhysRevD.111.094502} {\bibfield
  {journal} {\bibinfo  {journal} {Phys. Rev. D}\ }\textbf {\bibinfo {volume}
  {111}},\ \bibinfo {pages} {094502} (\bibinfo {year} {2025})},\ \Eprint
  {http://arxiv.org/abs/2502.17551} {arXiv:2502.17551 [hep-ph]} \BibitemShut
  {NoStop}%
\bibitem [{\citenamefont {Janik}\ \emph {et~al.}(2025)\citenamefont {Janik},
  \citenamefont {Nowak}, \citenamefont {Rams},\ and\ \citenamefont
  {Zahed}}]{Janik:2025bbz}%
  \BibitemOpen
  \bibfield  {author} {\bibinfo {author} {\bibfnamefont {Romuald~A.}\
  \bibnamefont {Janik}}, \bibinfo {author} {\bibfnamefont {Maciej~A.}\
  \bibnamefont {Nowak}}, \bibinfo {author} {\bibfnamefont {Marek~M.}\
  \bibnamefont {Rams}}, \ and\ \bibinfo {author} {\bibfnamefont {Ismail}\
  \bibnamefont {Zahed}},\ }\bibfield  {title} {\enquote {\bibinfo {title}
  {{Universality and emergent effective fluid from jets and string breaking in
  the massive Schwinger model using tensor networks}},}\ }\href@noop {} {\
  (\bibinfo {year} {2025})},\ \Eprint {http://arxiv.org/abs/2502.12901}
  {arXiv:2502.12901 [hep-ph]} \BibitemShut {NoStop}%
\bibitem [{\citenamefont {Shao}\ \emph {et~al.}(2025)\citenamefont {Shao},
  \citenamefont {Chen},\ and\ \citenamefont {Shi}}]{Shao:2025ygy}%
  \BibitemOpen
  \bibfield  {author} {\bibinfo {author} {\bibfnamefont {Haiyang}\ \bibnamefont
  {Shao}}, \bibinfo {author} {\bibfnamefont {Shile}\ \bibnamefont {Chen}}, \
  and\ \bibinfo {author} {\bibfnamefont {Shuzhe}\ \bibnamefont {Shi}},\
  }\bibfield  {title} {\enquote {\bibinfo {title} {{Onset of Bjorken Flow in
  Quantum Evolution of the Massive Schwinger Model}},}\ }\href@noop {} {\
  (\bibinfo {year} {2025})},\ \Eprint {http://arxiv.org/abs/2509.10855}
  {arXiv:2509.10855 [hep-ph]} \BibitemShut {NoStop}%
\bibitem [{\citenamefont {Ebner}\ \emph {et~al.}(2025)\citenamefont {Ebner},
  \citenamefont {M{\"u}ller}, \citenamefont {Sch{\"a}fer}, \citenamefont
  {Schmotzer}, \citenamefont {Seidl},\ and\ \citenamefont
  {Yao}}]{Ebner:2025pdm}%
  \BibitemOpen
  \bibfield  {author} {\bibinfo {author} {\bibfnamefont {Lukas}\ \bibnamefont
  {Ebner}}, \bibinfo {author} {\bibfnamefont {Berndt}\ \bibnamefont
  {M{\"u}ller}}, \bibinfo {author} {\bibfnamefont {Andreas}\ \bibnamefont
  {Sch{\"a}fer}}, \bibinfo {author} {\bibfnamefont {Leonhard}\ \bibnamefont
  {Schmotzer}}, \bibinfo {author} {\bibfnamefont {Clemens}\ \bibnamefont
  {Seidl}}, \ and\ \bibinfo {author} {\bibfnamefont {Xiaojun}\ \bibnamefont
  {Yao}},\ }\bibfield  {title} {\enquote {\bibinfo {title} {{The Magic Barrier
  before Thermalization}},}\ }\href@noop {} {\  (\bibinfo {year} {2025})},\
  \Eprint {http://arxiv.org/abs/2510.11681} {arXiv:2510.11681 [quant-ph]}
  \BibitemShut {NoStop}%
\bibitem [{\citenamefont {Artiaco}\ \emph {et~al.}(2025)\citenamefont
  {Artiaco}, \citenamefont {Barata},\ and\ \citenamefont
  {Rico}}]{Artiaco:2025qqq}%
  \BibitemOpen
  \bibfield  {author} {\bibinfo {author} {\bibfnamefont {Claudia}\ \bibnamefont
  {Artiaco}}, \bibinfo {author} {\bibfnamefont {Jo{\~a}o}\ \bibnamefont
  {Barata}}, \ and\ \bibinfo {author} {\bibfnamefont {Enrique}\ \bibnamefont
  {Rico}},\ }\bibfield  {title} {\enquote {\bibinfo {title}
  {{Out-of-Equilibrium Dynamics in a U(1) Lattice Gauge Theory via Local
  Information Flows: Scattering and String Breaking}},}\ }\href@noop {} {\
  (\bibinfo {year} {2025})},\ \Eprint {http://arxiv.org/abs/2510.16101}
  {arXiv:2510.16101 [quant-ph]} \BibitemShut {NoStop}%
\bibitem [{\citenamefont {Schwinger}(1962)}]{Schwinger:1962tp}%
  \BibitemOpen
  \bibfield  {author} {\bibinfo {author} {\bibfnamefont {Julian~S.}\
  \bibnamefont {Schwinger}},\ }\bibfield  {title} {\enquote {\bibinfo {title}
  {{Gauge Invariance and Mass. 2.}}}\ }\href {\doibase
  10.1103/PhysRev.128.2425} {\bibfield  {journal} {\bibinfo  {journal} {Phys.
  Rev.}\ }\textbf {\bibinfo {volume} {128}},\ \bibinfo {pages} {2425--2429}
  (\bibinfo {year} {1962})}\BibitemShut {NoStop}%
\bibitem [{\citenamefont {Coleman}(1976)}]{Coleman:1976uz}%
  \BibitemOpen
  \bibfield  {author} {\bibinfo {author} {\bibfnamefont {Sidney~R.}\
  \bibnamefont {Coleman}},\ }\bibfield  {title} {\enquote {\bibinfo {title}
  {{More About the Massive Schwinger Model}},}\ }\href {\doibase
  10.1016/0003-4916(76)90280-3} {\bibfield  {journal} {\bibinfo  {journal}
  {Annals Phys.}\ }\textbf {\bibinfo {volume} {101}},\ \bibinfo {pages} {239}
  (\bibinfo {year} {1976})}\BibitemShut {NoStop}%
\bibitem [{\citenamefont {Manton}(1985)}]{Manton:1985jm}%
  \BibitemOpen
  \bibfield  {author} {\bibinfo {author} {\bibfnamefont {N.~S.}\ \bibnamefont
  {Manton}},\ }\bibfield  {title} {\enquote {\bibinfo {title} {{The Schwinger
  Model and Its Axial Anomaly}},}\ }\href {\doibase
  10.1016/0003-4916(85)90199-X} {\bibfield  {journal} {\bibinfo  {journal}
  {Annals Phys.}\ }\textbf {\bibinfo {volume} {159}},\ \bibinfo {pages}
  {220--251} (\bibinfo {year} {1985})}\BibitemShut {NoStop}%
\bibitem [{\citenamefont {Chitambar}\ and\ \citenamefont
  {Gour}(2019)}]{chitambar2019quantum}%
  \BibitemOpen
  \bibfield  {author} {\bibinfo {author} {\bibfnamefont {Eric}\ \bibnamefont
  {Chitambar}}\ and\ \bibinfo {author} {\bibfnamefont {Gilad}\ \bibnamefont
  {Gour}},\ }\bibfield  {title} {\enquote {\bibinfo {title} {Quantum resource
  theories},}\ }\href {\doibase 10.1103/RevModPhys.91.025001} {\bibfield
  {journal} {\bibinfo  {journal} {Rev. Mod. Phys.}\ }\textbf {\bibinfo {volume}
  {91}},\ \bibinfo {pages} {025001} (\bibinfo {year} {2019})}\BibitemShut
  {NoStop}%
\bibitem [{\citenamefont {Aditya}\ \emph {et~al.}(2025)\citenamefont {Aditya},
  \citenamefont {Summer}, \citenamefont {Sierant},\ and\ \citenamefont
  {Turkeshi}}]{aditya2025mpemba}%
  \BibitemOpen
  \bibfield  {author} {\bibinfo {author} {\bibfnamefont {Sreemayee}\
  \bibnamefont {Aditya}}, \bibinfo {author} {\bibfnamefont {Alessandro}\
  \bibnamefont {Summer}}, \bibinfo {author} {\bibfnamefont {Piotr}\
  \bibnamefont {Sierant}}, \ and\ \bibinfo {author} {\bibfnamefont {Xhek}\
  \bibnamefont {Turkeshi}},\ }\href {https://arxiv.org/abs/2509.22176}
  {\enquote {\bibinfo {title} {Mpemba effects in quantum complexity},}\ }
  (\bibinfo {year} {2025}),\ \Eprint {http://arxiv.org/abs/2509.22176}
  {arXiv:2509.22176 [quant-ph]} \BibitemShut {NoStop}%
\bibitem [{\citenamefont {Summer}\ \emph {et~al.}(2025)\citenamefont {Summer},
  \citenamefont {Moroder}, \citenamefont {Bettmann}, \citenamefont {Turkeshi},
  \citenamefont {Marvian},\ and\ \citenamefont {Goold}}]{summer2025}%
  \BibitemOpen
  \bibfield  {author} {\bibinfo {author} {\bibfnamefont {Alessandro}\
  \bibnamefont {Summer}}, \bibinfo {author} {\bibfnamefont {Mattia}\
  \bibnamefont {Moroder}}, \bibinfo {author} {\bibfnamefont {Laetitia~P.}\
  \bibnamefont {Bettmann}}, \bibinfo {author} {\bibfnamefont {Xhek}\
  \bibnamefont {Turkeshi}}, \bibinfo {author} {\bibfnamefont {Iman}\
  \bibnamefont {Marvian}}, \ and\ \bibinfo {author} {\bibfnamefont {John}\
  \bibnamefont {Goold}},\ }\href {https://arxiv.org/abs/2507.16976} {\enquote
  {\bibinfo {title} {A resource theoretical unification of mpemba effects:
  classical and quantum},}\ } (\bibinfo {year} {2025}),\ \Eprint
  {http://arxiv.org/abs/2507.16976} {arXiv:2507.16976 [quant-ph]} \BibitemShut
  {NoStop}%
\bibitem [{\citenamefont {Capizzi}\ and\ \citenamefont
  {Mazzoni}(2023)}]{capizzi2023entanglement}%
  \BibitemOpen
  \bibfield  {author} {\bibinfo {author} {\bibfnamefont {Luca}\ \bibnamefont
  {Capizzi}}\ and\ \bibinfo {author} {\bibfnamefont {Michele}\ \bibnamefont
  {Mazzoni}},\ }\bibfield  {title} {\enquote {\bibinfo {title} {{Entanglement
  asymmetry in the ordered phase of many-body systems: the Ising field
  theory}},}\ }\href {\doibase 10.1007/JHEP12(2023)144} {\bibfield  {journal}
  {\bibinfo  {journal} {JHEP}\ }\textbf {\bibinfo {volume} {2023}},\ \bibinfo
  {pages} {144} (\bibinfo {year} {2023})}\BibitemShut {NoStop}%
\bibitem [{\citenamefont {Fossati}\ \emph {et~al.}(2024)\citenamefont
  {Fossati}, \citenamefont {Ares}, \citenamefont {Dubail},\ and\ \citenamefont
  {Calabrese}}]{fossati2024entanglement}%
  \BibitemOpen
  \bibfield  {author} {\bibinfo {author} {\bibfnamefont {Michele}\ \bibnamefont
  {Fossati}}, \bibinfo {author} {\bibfnamefont {Filiberto}\ \bibnamefont
  {Ares}}, \bibinfo {author} {\bibfnamefont {J{\'e}r{\^o}me}\ \bibnamefont
  {Dubail}}, \ and\ \bibinfo {author} {\bibfnamefont {Pasquale}\ \bibnamefont
  {Calabrese}},\ }\bibfield  {title} {\enquote {\bibinfo {title} {{Entanglement
  asymmetry in CFT and its relation to non-topological defects}},}\ }\href
  {\doibase 10.1007/JHEP05(2024)059} {\bibfield  {journal} {\bibinfo  {journal}
  {JHEP}\ }\textbf {\bibinfo {volume} {2024}},\ \bibinfo {pages} {059}
  (\bibinfo {year} {2024})}\BibitemShut {NoStop}%
\bibitem [{\citenamefont {Capizzi}\ and\ \citenamefont
  {Vitale}(2024)}]{capizzi2024universal}%
  \BibitemOpen
  \bibfield  {author} {\bibinfo {author} {\bibfnamefont {Luca}\ \bibnamefont
  {Capizzi}}\ and\ \bibinfo {author} {\bibfnamefont {Vittorio}\ \bibnamefont
  {Vitale}},\ }\bibfield  {title} {\enquote {\bibinfo {title} {A universal
  formula for the entanglement asymmetry of matrix product states},}\ }\href
  {\doibase 10.1088/1751-8121/ad8796} {\bibfield  {journal} {\bibinfo
  {journal} {J. Phys. A: Math. Theor.}\ }\textbf {\bibinfo {volume} {57}},\
  \bibinfo {pages} {45LT01} (\bibinfo {year} {2024})}\BibitemShut {NoStop}%
\bibitem [{\citenamefont {Chen}\ and\ \citenamefont {Chen}(2024)}]{chen2023}%
  \BibitemOpen
  \bibfield  {author} {\bibinfo {author} {\bibfnamefont {Miao}\ \bibnamefont
  {Chen}}\ and\ \bibinfo {author} {\bibfnamefont {Hui-Huang}\ \bibnamefont
  {Chen}},\ }\bibfield  {title} {\enquote {\bibinfo {title} {R\'enyi
  entanglement asymmetry in ($1+1$)-dimensional conformal field theories},}\
  }\href {\doibase 10.1103/PhysRevD.109.065009} {\bibfield  {journal} {\bibinfo
   {journal} {Phys. Rev. D}\ }\textbf {\bibinfo {volume} {109}},\ \bibinfo
  {pages} {065009} (\bibinfo {year} {2024})}\BibitemShut {NoStop}%
\bibitem [{\citenamefont {Ares}\ \emph {et~al.}(2024)\citenamefont {Ares},
  \citenamefont {Murciano}, \citenamefont {Piroli},\ and\ \citenamefont
  {Calabrese}}]{ares2024entanglement}%
  \BibitemOpen
  \bibfield  {author} {\bibinfo {author} {\bibfnamefont {Filiberto}\
  \bibnamefont {Ares}}, \bibinfo {author} {\bibfnamefont {Sara}\ \bibnamefont
  {Murciano}}, \bibinfo {author} {\bibfnamefont {Lorenzo}\ \bibnamefont
  {Piroli}}, \ and\ \bibinfo {author} {\bibfnamefont {Pasquale}\ \bibnamefont
  {Calabrese}},\ }\bibfield  {title} {\enquote {\bibinfo {title} {Entanglement
  asymmetry study of black hole radiation},}\ }\href {\doibase
  10.1103/PhysRevD.110.L061901} {\bibfield  {journal} {\bibinfo  {journal}
  {Phys. Rev. D}\ }\textbf {\bibinfo {volume} {110}},\ \bibinfo {pages}
  {L061901} (\bibinfo {year} {2024})}\BibitemShut {NoStop}%
\bibitem [{\citenamefont {Russotto}\ \emph {et~al.}(2024)\citenamefont
  {Russotto}, \citenamefont {Ares},\ and\ \citenamefont
  {Calabrese}}]{russotto2024non}%
  \BibitemOpen
  \bibfield  {author} {\bibinfo {author} {\bibfnamefont {Angelo}\ \bibnamefont
  {Russotto}}, \bibinfo {author} {\bibfnamefont {Filiberto}\ \bibnamefont
  {Ares}}, \ and\ \bibinfo {author} {\bibfnamefont {Pasquale}\ \bibnamefont
  {Calabrese}},\ }\bibfield  {title} {\enquote {\bibinfo {title} {{Non-Abelian
  entanglement asymmetry in random states}},}\ }\href
  {https://arxiv.org/abs/2411.13337} {\bibfield  {journal} {\bibinfo  {journal}
  {arXiv:2411.13337}\ } (\bibinfo {year} {2024})}\BibitemShut {NoStop}%
\bibitem [{\citenamefont {Lastres}\ \emph {et~al.}(2024)\citenamefont
  {Lastres}, \citenamefont {Murciano}, \citenamefont {Ares},\ and\
  \citenamefont {Calabrese}}]{lastres2024entanglement}%
  \BibitemOpen
  \bibfield  {author} {\bibinfo {author} {\bibfnamefont {Marco}\ \bibnamefont
  {Lastres}}, \bibinfo {author} {\bibfnamefont {Sara}\ \bibnamefont
  {Murciano}}, \bibinfo {author} {\bibfnamefont {Filiberto}\ \bibnamefont
  {Ares}}, \ and\ \bibinfo {author} {\bibfnamefont {Pasquale}\ \bibnamefont
  {Calabrese}},\ }\bibfield  {title} {\enquote {\bibinfo {title} {{Entanglement
  asymmetry in the critical XXZ spin chain}},}\ }\href
  {https://arxiv.org/abs/2407.06427} {\bibfield  {journal} {\bibinfo  {journal}
  {arXiv:2407.06427}\ } (\bibinfo {year} {2024})}\BibitemShut {NoStop}%
\bibitem [{\citenamefont {Fossati}\ \emph {et~al.}(2025)\citenamefont
  {Fossati}, \citenamefont {Rylands},\ and\ \citenamefont
  {Calabrese}}]{fossati2024}%
  \BibitemOpen
  \bibfield  {author} {\bibinfo {author} {\bibfnamefont {Michele}\ \bibnamefont
  {Fossati}}, \bibinfo {author} {\bibfnamefont {Colin}\ \bibnamefont
  {Rylands}}, \ and\ \bibinfo {author} {\bibfnamefont {Pasquale}\ \bibnamefont
  {Calabrese}},\ }\bibfield  {title} {\enquote {\bibinfo {title} {{Entanglement
  asymmetry in CFT with boundary symmetry breaking}},}\ }\href {\doibase
  10.1007/JHEP06(2025)089} {\bibfield  {journal} {\bibinfo  {journal} {JHEP}\
  }\textbf {\bibinfo {volume} {06}},\ \bibinfo {pages} {089} (\bibinfo {year}
  {2025})},\ \Eprint {http://arxiv.org/abs/2411.10244} {arXiv:2411.10244
  [hep-th]} \BibitemShut {NoStop}%
\bibitem [{\citenamefont {Kusuki}\ \emph {et~al.}(2025)\citenamefont {Kusuki},
  \citenamefont {Murciano}, \citenamefont {Ooguri},\ and\ \citenamefont
  {Pal}}]{kusuki2024}%
  \BibitemOpen
  \bibfield  {author} {\bibinfo {author} {\bibfnamefont {Yuya}\ \bibnamefont
  {Kusuki}}, \bibinfo {author} {\bibfnamefont {Sara}\ \bibnamefont {Murciano}},
  \bibinfo {author} {\bibfnamefont {Hirosi}\ \bibnamefont {Ooguri}}, \ and\
  \bibinfo {author} {\bibfnamefont {Sridip}\ \bibnamefont {Pal}},\ }\bibfield
  {title} {\enquote {\bibinfo {title} {{Entanglement asymmetry and symmetry
  defects in boundary conformal field theory}},}\ }\href {\doibase
  10.1007/JHEP01(2025)057} {\bibfield  {journal} {\bibinfo  {journal} {JHEP}\
  }\textbf {\bibinfo {volume} {01}},\ \bibinfo {pages} {057 (2025)} (\bibinfo
  {year} {2025})}\BibitemShut {NoStop}%
\bibitem [{\citenamefont {Chen}(2024)}]{chen2024}%
  \BibitemOpen
  \bibfield  {author} {\bibinfo {author} {\bibfnamefont {Hui-Huang}\
  \bibnamefont {Chen}},\ }\bibfield  {title} {\enquote {\bibinfo {title}
  {{Entanglement asymmetry in the Hayden-Preskill protocol}},}\ }\href
  {https://arxiv.org/abs/2411.17695} {\bibfield  {journal} {\bibinfo  {journal}
  {arXiv:2411.17695}\ } (\bibinfo {year} {2024})}\BibitemShut {NoStop}%
\bibitem [{\citenamefont {Russotto}\ \emph {et~al.}(2025)\citenamefont
  {Russotto}, \citenamefont {Ares},\ and\ \citenamefont
  {Calabrese}}]{russotto2025}%
  \BibitemOpen
  \bibfield  {author} {\bibinfo {author} {\bibfnamefont {Angelo}\ \bibnamefont
  {Russotto}}, \bibinfo {author} {\bibfnamefont {Filiberto}\ \bibnamefont
  {Ares}}, \ and\ \bibinfo {author} {\bibfnamefont {Pasquale}\ \bibnamefont
  {Calabrese}},\ }\bibfield  {title} {\enquote {\bibinfo {title} {Symmetry
  breaking in chaotic many-body quantum systems at finite temperature},}\
  }\href {\doibase 10.1103/kppn-3272} {\bibfield  {journal} {\bibinfo
  {journal} {Phys. Rev. E}\ }\textbf {\bibinfo {volume} {112}},\ \bibinfo
  {pages} {L032101} (\bibinfo {year} {2025})}\BibitemShut {NoStop}%
\bibitem [{\citenamefont {Mazzoni}\ \emph {et~al.}(2025)\citenamefont
  {Mazzoni}, \citenamefont {Capizzi},\ and\ \citenamefont
  {Piroli}}]{Mazzoni:2025otu}%
  \BibitemOpen
  \bibfield  {author} {\bibinfo {author} {\bibfnamefont {Michele}\ \bibnamefont
  {Mazzoni}}, \bibinfo {author} {\bibfnamefont {Luca}\ \bibnamefont {Capizzi}},
  \ and\ \bibinfo {author} {\bibfnamefont {Lorenzo}\ \bibnamefont {Piroli}},\
  }\bibfield  {title} {\enquote {\bibinfo {title} {{Breaking global symmetries
  with locality-preserving operations}},}\ }\href@noop {} {\  (\bibinfo {year}
  {2025})},\ \Eprint {http://arxiv.org/abs/2508.15892} {arXiv:2508.15892
  [quant-ph]} \BibitemShut {NoStop}%
\bibitem [{\citenamefont {Lamas}\ \emph {et~al.}(2025)\citenamefont {Lamas},
  \citenamefont {Gliozzi},\ and\ \citenamefont {Hughes}}]{lamas2025}%
  \BibitemOpen
  \bibfield  {author} {\bibinfo {author} {\bibfnamefont {Amanda~Gatto}\
  \bibnamefont {Lamas}}, \bibinfo {author} {\bibfnamefont {Jacopo}\
  \bibnamefont {Gliozzi}}, \ and\ \bibinfo {author} {\bibfnamefont {Taylor~L.}\
  \bibnamefont {Hughes}},\ }\href {https://arxiv.org/abs/2510.03967} {\enquote
  {\bibinfo {title} {Higher-form entanglement asymmetry and topological
  order},}\ } (\bibinfo {year} {2025}),\ \Eprint
  {http://arxiv.org/abs/2510.03967} {arXiv:2510.03967 [cond-mat.str-el]}
  \BibitemShut {NoStop}%
\bibitem [{\citenamefont {Benini}\ \emph {et~al.}(2025)\citenamefont {Benini},
  \citenamefont {Calabrese}, \citenamefont {Fossati}, \citenamefont {Singh},\
  and\ \citenamefont {Venuti}}]{Benini:2025lav}%
  \BibitemOpen
  \bibfield  {author} {\bibinfo {author} {\bibfnamefont {Francesco}\
  \bibnamefont {Benini}}, \bibinfo {author} {\bibfnamefont {Pasquale}\
  \bibnamefont {Calabrese}}, \bibinfo {author} {\bibfnamefont {Michele}\
  \bibnamefont {Fossati}}, \bibinfo {author} {\bibfnamefont {Amartya~Harsh}\
  \bibnamefont {Singh}}, \ and\ \bibinfo {author} {\bibfnamefont {Marco}\
  \bibnamefont {Venuti}},\ }\bibfield  {title} {\enquote {\bibinfo {title}
  {{Entanglement Asymmetry for Higher and Noninvertible Symmetries}},}\
  }\href@noop {} {\  (\bibinfo {year} {2025})},\ \Eprint
  {http://arxiv.org/abs/2509.16311} {arXiv:2509.16311 [hep-th]} \BibitemShut
  {NoStop}%
\bibitem [{\citenamefont {Ahmad}\ \emph {et~al.}(2025)\citenamefont {Ahmad},
  \citenamefont {Klinger},\ and\ \citenamefont {Wang}}]{ahmad2025}%
  \BibitemOpen
  \bibfield  {author} {\bibinfo {author} {\bibfnamefont {Shadi~Ali}\
  \bibnamefont {Ahmad}}, \bibinfo {author} {\bibfnamefont {Marc~S.}\
  \bibnamefont {Klinger}}, \ and\ \bibinfo {author} {\bibfnamefont {Yifan}\
  \bibnamefont {Wang}},\ }\href {https://arxiv.org/abs/2509.18072} {\enquote
  {\bibinfo {title} {The many faces of non-invertible symmetries},}\ }
  (\bibinfo {year} {2025}),\ \Eprint {http://arxiv.org/abs/2509.18072}
  {arXiv:2509.18072 [hep-th]} \BibitemShut {NoStop}%
\bibitem [{\citenamefont {Ares}\ \emph
  {et~al.}(2025{\natexlab{a}})\citenamefont {Ares}, \citenamefont {Calabrese},\
  and\ \citenamefont {Murciano}}]{review}%
  \BibitemOpen
  \bibfield  {author} {\bibinfo {author} {\bibfnamefont {Filiberto}\
  \bibnamefont {Ares}}, \bibinfo {author} {\bibfnamefont {Pasquale}\
  \bibnamefont {Calabrese}}, \ and\ \bibinfo {author} {\bibfnamefont {Sara}\
  \bibnamefont {Murciano}},\ }\bibfield  {title} {\enquote {\bibinfo {title}
  {{The quantum Mpemba effects}},}\ }\href {\doibase
  10.1038/s42254-025-00838-0} {\bibfield  {journal} {\bibinfo  {journal}
  {Nature Rev. Phys.}\ }\textbf {\bibinfo {volume} {7}},\ \bibinfo {pages}
  {451--460} (\bibinfo {year} {2025}{\natexlab{a}})},\ \Eprint
  {http://arxiv.org/abs/2502.08087} {arXiv:2502.08087 [cond-mat.stat-mech]}
  \BibitemShut {NoStop}%
\bibitem [{\citenamefont {Teza}\ \emph {et~al.}(2025)\citenamefont {Teza},
  \citenamefont {Bechhoefer}, \citenamefont {Lasanta}, \citenamefont {Raz},\
  and\ \citenamefont {Vucelja}}]{teza2025}%
  \BibitemOpen
  \bibfield  {author} {\bibinfo {author} {\bibfnamefont {Gianluca}\
  \bibnamefont {Teza}}, \bibinfo {author} {\bibfnamefont {John}\ \bibnamefont
  {Bechhoefer}}, \bibinfo {author} {\bibfnamefont {Antonio}\ \bibnamefont
  {Lasanta}}, \bibinfo {author} {\bibfnamefont {Oren}\ \bibnamefont {Raz}}, \
  and\ \bibinfo {author} {\bibfnamefont {Marija}\ \bibnamefont {Vucelja}},\
  }\href {https://arxiv.org/abs/2502.01758} {\enquote {\bibinfo {title}
  {Speedups in nonequilibrium thermal relaxation: Mpemba and related
  effects},}\ } (\bibinfo {year} {2025}),\ \Eprint
  {http://arxiv.org/abs/2502.01758} {arXiv:2502.01758 [cond-mat.stat-mech]}
  \BibitemShut {NoStop}%
\bibitem [{\citenamefont {Murciano}\ \emph {et~al.}(2024)\citenamefont
  {Murciano}, \citenamefont {Ares}, \citenamefont {Klich},\ and\ \citenamefont
  {Calabrese}}]{murciano2024}%
  \BibitemOpen
  \bibfield  {author} {\bibinfo {author} {\bibfnamefont {Sara}\ \bibnamefont
  {Murciano}}, \bibinfo {author} {\bibfnamefont {Filiberto}\ \bibnamefont
  {Ares}}, \bibinfo {author} {\bibfnamefont {Israel}\ \bibnamefont {Klich}}, \
  and\ \bibinfo {author} {\bibfnamefont {Pasquale}\ \bibnamefont {Calabrese}},\
  }\bibfield  {title} {\enquote {\bibinfo {title} {Entanglement asymmetry and
  quantum mpemba effect in the xy spin chain},}\ }\href {\doibase
  10.1088/1742-5468/ad17b4} {\bibfield  {journal} {\bibinfo  {journal} {Journal
  of Statistical Mechanics: Theory and Experiment}\ }\textbf {\bibinfo {volume}
  {2024}},\ \bibinfo {pages} {013103} (\bibinfo {year} {2024})}\BibitemShut
  {NoStop}%
\bibitem [{\citenamefont {Sachs}\ and\ \citenamefont
  {Wipf}(1992)}]{Sachs:1991en}%
  \BibitemOpen
  \bibfield  {author} {\bibinfo {author} {\bibfnamefont {Ivo}\ \bibnamefont
  {Sachs}}\ and\ \bibinfo {author} {\bibfnamefont {Andreas}\ \bibnamefont
  {Wipf}},\ }\bibfield  {title} {\enquote {\bibinfo {title} {{Finite
  temperature Schwinger model}},}\ }\href@noop {} {\bibfield  {journal}
  {\bibinfo  {journal} {Helv. Phys. Acta}\ }\textbf {\bibinfo {volume} {65}},\
  \bibinfo {pages} {652--678} (\bibinfo {year} {1992})},\ \Eprint
  {http://arxiv.org/abs/1005.1822} {arXiv:1005.1822 [hep-th]} \BibitemShut
  {NoStop}%
\bibitem [{\citenamefont {Polikarpov}\ and\ \citenamefont
  {Buividovich}(2008)}]{Polikarpov:2008tg}%
  \BibitemOpen
  \bibfield  {author} {\bibinfo {author} {\bibfnamefont {M.~I.}\ \bibnamefont
  {Polikarpov}}\ and\ \bibinfo {author} {\bibfnamefont {P.~V.}\ \bibnamefont
  {Buividovich}},\ }\bibfield  {title} {\enquote {\bibinfo {title} {{Z2
  electric strings and center vortices in SU(2) lattice gauge theory}},}\ }in\
  \href@noop {} {\emph {\bibinfo {booktitle} {{13th Lomonosov Conference on
  Elementary Particle Physics}}}}\ (\bibinfo {year} {2008})\ pp.\ \bibinfo
  {pages} {378--385},\ \Eprint {http://arxiv.org/abs/0801.0262}
  {arXiv:0801.0262 [hep-lat]} \BibitemShut {NoStop}%
\bibitem [{\citenamefont {Bulgarelli}\ and\ \citenamefont
  {Panero}(2023)}]{Bulgarelli:2023ofi}%
  \BibitemOpen
  \bibfield  {author} {\bibinfo {author} {\bibfnamefont {Andrea}\ \bibnamefont
  {Bulgarelli}}\ and\ \bibinfo {author} {\bibfnamefont {Marco}\ \bibnamefont
  {Panero}},\ }\bibfield  {title} {\enquote {\bibinfo {title} {{Entanglement
  entropy from non-equilibrium Monte Carlo simulations}},}\ }\href {\doibase
  10.1007/JHEP06(2023)030} {\bibfield  {journal} {\bibinfo  {journal} {JHEP}\
  }\textbf {\bibinfo {volume} {06}},\ \bibinfo {pages} {030} (\bibinfo {year}
  {2023})},\ \Eprint {http://arxiv.org/abs/2304.03311} {arXiv:2304.03311
  [quant-ph]} \BibitemShut {NoStop}%
\bibitem [{\citenamefont {Bulgarelli}\ and\ \citenamefont
  {Panero}(2024)}]{Bulgarelli:2024onj}%
  \BibitemOpen
  \bibfield  {author} {\bibinfo {author} {\bibfnamefont {Andrea}\ \bibnamefont
  {Bulgarelli}}\ and\ \bibinfo {author} {\bibfnamefont {Marco}\ \bibnamefont
  {Panero}},\ }\bibfield  {title} {\enquote {\bibinfo {title} {{Duality
  transformations and the entanglement entropy of gauge theories}},}\ }\href
  {\doibase 10.1007/JHEP06(2024)041} {\bibfield  {journal} {\bibinfo  {journal}
  {JHEP}\ }\textbf {\bibinfo {volume} {06}},\ \bibinfo {pages} {041} (\bibinfo
  {year} {2024})},\ \Eprint {http://arxiv.org/abs/2404.01987} {arXiv:2404.01987
  [quant-ph]} \BibitemShut {NoStop}%
\bibitem [{\citenamefont {Amorosso}\ \emph {et~al.}(2024)\citenamefont
  {Amorosso}, \citenamefont {Syritsyn},\ and\ \citenamefont
  {Venugopalan}}]{Amorosso:2024leg}%
  \BibitemOpen
  \bibfield  {author} {\bibinfo {author} {\bibfnamefont {Rocco}\ \bibnamefont
  {Amorosso}}, \bibinfo {author} {\bibfnamefont {Sergey}\ \bibnamefont
  {Syritsyn}}, \ and\ \bibinfo {author} {\bibfnamefont {Raju}\ \bibnamefont
  {Venugopalan}},\ }\bibfield  {title} {\enquote {\bibinfo {title}
  {{Entanglement entropy of a color flux tube in (2+1)D Yang-Mills theory}},}\
  }\href {\doibase 10.1007/JHEP12(2024)177} {\bibfield  {journal} {\bibinfo
  {journal} {JHEP}\ }\textbf {\bibinfo {volume} {12}},\ \bibinfo {pages} {177}
  (\bibinfo {year} {2024})},\ \Eprint {http://arxiv.org/abs/2410.00112}
  {arXiv:2410.00112 [hep-lat]} \BibitemShut {NoStop}%
\bibitem [{\citenamefont {Amorosso}\ \emph {et~al.}(2025)\citenamefont
  {Amorosso}, \citenamefont {Syritsyn},\ and\ \citenamefont
  {Venugopalan}}]{Amorosso:2024glf}%
  \BibitemOpen
  \bibfield  {author} {\bibinfo {author} {\bibfnamefont {Rocco}\ \bibnamefont
  {Amorosso}}, \bibinfo {author} {\bibfnamefont {Sergey}\ \bibnamefont
  {Syritsyn}}, \ and\ \bibinfo {author} {\bibfnamefont {Raju}\ \bibnamefont
  {Venugopalan}},\ }\bibfield  {title} {\enquote {\bibinfo {title}
  {{Entanglement entropy of a color flux tube in (1+1)D Yang{\textendash}Mills
  theory}},}\ }\href {\doibase 10.1016/j.physletb.2025.139806} {\bibfield
  {journal} {\bibinfo  {journal} {Phys. Lett. B}\ }\textbf {\bibinfo {volume}
  {868}},\ \bibinfo {pages} {139806} (\bibinfo {year} {2025})},\ \Eprint
  {http://arxiv.org/abs/2411.12818} {arXiv:2411.12818 [hep-lat]} \BibitemShut
  {NoStop}%
\bibitem [{\citenamefont {Sala}\ \emph {et~al.}(2018)\citenamefont {Sala},
  \citenamefont {Shi}, \citenamefont {K\"uhn}, \citenamefont {Ba\~nuls},
  \citenamefont {Demler},\ and\ \citenamefont {Cirac}}]{sala2018}%
  \BibitemOpen
  \bibfield  {author} {\bibinfo {author} {\bibfnamefont {P.}~\bibnamefont
  {Sala}}, \bibinfo {author} {\bibfnamefont {T.}~\bibnamefont {Shi}}, \bibinfo
  {author} {\bibfnamefont {S.}~\bibnamefont {K\"uhn}}, \bibinfo {author}
  {\bibfnamefont {M.~C.}\ \bibnamefont {Ba\~nuls}}, \bibinfo {author}
  {\bibfnamefont {E.}~\bibnamefont {Demler}}, \ and\ \bibinfo {author}
  {\bibfnamefont {J.~I.}\ \bibnamefont {Cirac}},\ }\bibfield  {title} {\enquote
  {\bibinfo {title} {Variational study of u(1) and su(2) lattice gauge theories
  with gaussian states in $1+1$ dimensions},}\ }\href {\doibase
  10.1103/PhysRevD.98.034505} {\bibfield  {journal} {\bibinfo  {journal} {Phys.
  Rev. D}\ }\textbf {\bibinfo {volume} {98}},\ \bibinfo {pages} {034505}
  (\bibinfo {year} {2018})}\BibitemShut {NoStop}%
\bibitem [{\citenamefont {Choi}\ \emph {et~al.}(2022)\citenamefont {Choi},
  \citenamefont {Lam},\ and\ \citenamefont {Shao}}]{Choi:2022jqy}%
  \BibitemOpen
  \bibfield  {author} {\bibinfo {author} {\bibfnamefont {Yichul}\ \bibnamefont
  {Choi}}, \bibinfo {author} {\bibfnamefont {Ho~Tat}\ \bibnamefont {Lam}}, \
  and\ \bibinfo {author} {\bibfnamefont {Shu-Heng}\ \bibnamefont {Shao}},\
  }\bibfield  {title} {\enquote {\bibinfo {title} {{Noninvertible Global
  Symmetries in the Standard Model}},}\ }\href {\doibase
  10.1103/PhysRevLett.129.161601} {\bibfield  {journal} {\bibinfo  {journal}
  {Phys. Rev. Lett.}\ }\textbf {\bibinfo {volume} {129}},\ \bibinfo {pages}
  {161601} (\bibinfo {year} {2022})},\ \Eprint
  {http://arxiv.org/abs/2205.05086} {arXiv:2205.05086 [hep-th]} \BibitemShut
  {NoStop}%
\bibitem [{\citenamefont {Cordova}\ and\ \citenamefont
  {Ohmori}(2023)}]{Cordova:2022ieu}%
  \BibitemOpen
  \bibfield  {author} {\bibinfo {author} {\bibfnamefont {Clay}\ \bibnamefont
  {Cordova}}\ and\ \bibinfo {author} {\bibfnamefont {Kantaro}\ \bibnamefont
  {Ohmori}},\ }\bibfield  {title} {\enquote {\bibinfo {title} {{Noninvertible
  Chiral Symmetry and Exponential Hierarchies}},}\ }\href {\doibase
  10.1103/PhysRevX.13.011034} {\bibfield  {journal} {\bibinfo  {journal} {Phys.
  Rev. X}\ }\textbf {\bibinfo {volume} {13}},\ \bibinfo {pages} {011034}
  (\bibinfo {year} {2023})},\ \Eprint {http://arxiv.org/abs/2205.06243}
  {arXiv:2205.06243 [hep-th]} \BibitemShut {NoStop}%
\bibitem [{\citenamefont {Garc{\'\i}a~Etxebarria}\ and\ \citenamefont
  {Iqbal}(2023)}]{GarciaEtxebarria:2022jky}%
  \BibitemOpen
  \bibfield  {author} {\bibinfo {author} {\bibfnamefont {I{\~n}aki}\
  \bibnamefont {Garc{\'\i}a~Etxebarria}}\ and\ \bibinfo {author} {\bibfnamefont
  {Nabil}\ \bibnamefont {Iqbal}},\ }\bibfield  {title} {\enquote {\bibinfo
  {title} {{A Goldstone theorem for continuous non-invertible symmetries}},}\
  }\href {\doibase 10.1007/JHEP09(2023)145} {\bibfield  {journal} {\bibinfo
  {journal} {JHEP}\ }\textbf {\bibinfo {volume} {09}},\ \bibinfo {pages} {145}
  (\bibinfo {year} {2023})},\ \Eprint {http://arxiv.org/abs/2211.09570}
  {arXiv:2211.09570 [hep-th]} \BibitemShut {NoStop}%
\bibitem [{\citenamefont {Feynman}\ and\ \citenamefont
  {Hibbs}(2010)}]{Feynman2010-gl}%
  \BibitemOpen
  \bibfield  {author} {\bibinfo {author} {\bibfnamefont {Richard~P}\
  \bibnamefont {Feynman}}\ and\ \bibinfo {author} {\bibfnamefont {A~R}\
  \bibnamefont {Hibbs}},\ }\href@noop {} {\emph {\bibinfo {title} {Quantum
  mechanics and path integrals}}},\ Dover Books on Physics\ (\bibinfo
  {publisher} {Dover Publications},\ \bibinfo {address} {Mineola, NY},\
  \bibinfo {year} {2010})\BibitemShut {NoStop}%
\bibitem [{\citenamefont {Ares}\ \emph
  {et~al.}(2025{\natexlab{b}})\citenamefont {Ares}, \citenamefont {Vitale},\
  and\ \citenamefont {Murciano}}]{ares2024}%
  \BibitemOpen
  \bibfield  {author} {\bibinfo {author} {\bibfnamefont {Filiberto}\
  \bibnamefont {Ares}}, \bibinfo {author} {\bibfnamefont {Vittorio}\
  \bibnamefont {Vitale}}, \ and\ \bibinfo {author} {\bibfnamefont {Sara}\
  \bibnamefont {Murciano}},\ }\bibfield  {title} {\enquote {\bibinfo {title}
  {Quantum mpemba effect in free-fermionic mixed states},}\ }\href {\doibase
  10.1103/PhysRevB.111.104312} {\bibfield  {journal} {\bibinfo  {journal}
  {Phys. Rev. B}\ }\textbf {\bibinfo {volume} {111}},\ \bibinfo {pages}
  {104312} (\bibinfo {year} {2025}{\natexlab{b}})}\BibitemShut {NoStop}%
\end{thebibliography}
\end{document}